\documentclass[aps,pra,amsmath,amssymb, twocolumn,showpacs,longbibliography]{revtex4-1}

\usepackage{braket}
\usepackage{graphicx}

\usepackage{subfigure}
\usepackage[colorlinks, linkcolor=blue, urlcolor=blue, citecolor=blue]{hyperref} % Ceci est la bonne version pour avoir les liens DOI en biblio.
\usepackage{multirow}
\usepackage[normalem]{ulem}
\begin{document}

\title{Electron energy loss spectroscopy  of bulk gold 
with ultrasoft pseudopotentials and the Liouville-Lanczos method }
\author{Oleksandr Motornyi$^{1}$}
\email{motorny.sasha@gmail.com}
\author{Nathalie Vast$^{1}$}
\email{nathalie.vast@polytechnique.edu}
\author{Iurii Timrov$^{2}$}
\author{Oscar Baseggio$^{3}$}
\author{Stefano Baroni$^{3,4}$}
\author{Andrea Dal Corso$^{3,4}$}
\affiliation{$^{1}$ Laboratoire des Solides Irradi\'{e}s, CEA/DRF/IRAMIS, \'Ecole Polytechnique, CNRS,  Institut Polytechnique de Paris, 91128 Palaiseau c\'edex, France}
\affiliation{$^{2}$Theory and Simulation of Materials (THEOS), and National Centre for Computational Design and Discovery of Novel Materials (MARVEL), \'Ecole Polytechnique F\'ed\'erale de Lausanne (EPFL), CH-1015 Lausanne, Switzerland}
\affiliation{$^{3}$ Scuola Internazionale Superiore di Studi Avanzati (SISSA), Via Bonomea 265, IT-34136 Trieste, Italy}
\affiliation{$^{4}$ CNR-IOM DEMOCRITOS, Via Bonomea 265, IT-34136 Trieste, Italy}

\date{\today}

\begin{abstract}
The implementation of ultrasoft pseudopotentials into time-dependent density-functional perturbation theory is detailed for both the Sternheimer approach and the Liouville-Lanczos (LL) method, and 
equations are presented in the scalar relativistic approximation for periodic solids with finite momentum transfer {\bf q}. 
The LL method is applied to calculations of the electron energy loss (EEL) spectrum of face-centered cubic bulk Au both at vanishing and finite {\bf q}. 
Our study reveals the richness of the physics underlying the various contributions to the density fluctuation in gold. In particular, our calculations suggest the existence in gold of two quasi-separate $5d$ and $6s$  electron gasses, each one oscillating with its own frequency at resp. 5.1~eV and 10.2~eV.
We find that the contribution near 2.2~eV comes from $5d \rightarrow 6s$ interband transitions modified by the intraband 
contribution to the real part of the dielectric function, 
which we call a mixed excitation. 
%\Vchange{Finally,  the influence of the inclusion of the $5s$ and $5p$ semicore states into the valence region is discussed.}{}
\end{abstract}

\pacs{Condensed-Matter Physics, DFT development, EELS, plasmons, noble metals}
\maketitle

\section{Introduction}
\label{Intro}

Experiments that probe the dielectric function of finite systems are usefully complemented with \textit{ab initio} calculations based on time-dependent density-functional theory (TDDFT)~\cite{Runge:1984,Gross:1985,Gross:1996}, 
and sometimes also with many-body perturbation theory, for systems in which excitonic effects are important~\cite{Olevano:2001,Onida:2002, Anderson:2019,Vorwek:2019} and/or plasmon-phonon interaction is strong, leading to the presence of satellites in the spectrum~\cite{Kas:2014,Guzzo:2014,Nery:2016,Nery:2018}. 

Various advances have been made in the implementations of TDDFT. On the one hand, fully first-principles nonequilibrium simulations based on real-time time-dependent density functional theory (RT-TDDFT) are now accessible~\cite{Yost:2017,Tancogne-Dejean:2017,Miyamoto:2018}. 
On the other hand, the use of perturbation techniques has allowed to progress towards an efficient treatment of the linear response
of an electronic system to external perturbations within time-dependent (TD) density-functional perturbation theory (DFPT), based on the solution of the Sternheimer equation~\cite{Baroni:2001,Motornyi:2018}. 
Moreover, the method based on the Lanczos recursion method to solve the quantum Liouville equation, called the Liouville-Lanczos (LL) method, has allowed to speed-up the efficiency of calculations of the TDDFPT spectra, avoiding to solve the Sternheimer equations at each frequency~\cite{Walker:2006,Rocca:2008,Baroni:2012}.  

The LL method has been applied to the computation of optical spectra of molecular systems of unprecedented large size~\cite{Rocca:2009,Malcioglu:2011,Ghosh:2011,Gebauer:2013}. 
Since then, it has been used in the framework of many-body perturbation theory to capture
the electron-hole interaction in the Bethe-Salpeter equation~\cite{Rocca:2010,Rocca:2012}.
The LL approach to TDDFPT has also been extended to periodic solids and to finite values of the transferred momentum within a norm-conserving pseudopotential (NC-PP) framework to model plasmons~\cite{Timrov:2013,Timrov:2013b,Timrov:2013d,Timrov:2015}. The aim was to provide a valuable and computationally efficient theoretical tool to complement experiments that probe the dynamical structure factor, as measured in inelastic x-ray scattering (IXS) or the inverse dielectric function, as measured in electron energy loss spectroscopy~(EELS) experiments. Moreover, very recently, the LL method has been also generalized to model magnons in magnetic periodic solids~\cite{Gorni:2018}. 

In the present work we discuss the generalization of the LL method for EELS to ultrasoft pseudopotentials (US-PPs) with an  application to bulk gold. 
The objective is to have a tool to  investigate plasmons in systems of large size like surfaces with steps (vicinal surfaces)\cite{Motornyi:2018b}. 
As the efficiency of the treatment of large surfaces in the slab approach is heavily linked to the size of the plane wave basis set, a significant speed up in the calculations   
can be obtained when the kinetic energy cut-off of the plane waves is reduced. To this end, US-PPs have been developed~\cite{Vanderbilt:1990} about three decades ago to deal with
electronic states localized near the nucleus of an atom.
Actually the lift of the norm conservation constraint for the pseudo-wavefunctions, 
and the use of several reference energies for each angular momentum, with the multi-projector scheme inherent to the US-PP formalism,  
lead to accurate PPs even with a small kinetic energy cut-off for the plane waves. 
Numerous developments have extended the use of US-PPs in DFPT~\cite{Baroni:2001}, for instance for lattice dynamics~\cite{Dalcorso:1997,Dalcorso:2001}, electric field perturbations~\cite{Tobik:2004}, and TDDFPT for optical absorption of molecules~\cite{Walker:2007}. Most of the work has been done for 
the scalar relativistic (SR) US-PP scheme, 
but both DFT~\cite{Dalcorso:2005} and DFPT for lattice dynamics~\cite{Dalcorso:2007} have been generalized to the fully relativistic US-PPs scheme, including 
spin-orbit interaction in the solution of the Sternheimer linear system. 

The LL approach is similar to the Sternheimer method, in the sense that it does not need to perform expensive summations over empty states. However, in contrast to the Sternheimer method, there is no need to perform computations at each value of the excitation frequency: this is possible thanks to the use of the recursive Lanczos algorithm, which allows to obtain the charge-density susceptibility on an arbitrarily wide energy range with just one Lanczos chain. Due to the efficiency of the LL approach, it is used in this work and extended to the use of US-PPs. 
The LL method is then applied to the calculation of the electron energy loss (EEL)
spectrum of fcc-Au. 
%\Vchange{with and without $5s$ and $5p$ semicore states using US-PPs.}{}
Revisiting the EEL spectrum for $\mathbf{q} \rightarrow 0$, we provide a complete characterization and interpretation of all peaks and moreover we provide some more insights into the  understanding of the origin of certain excitations. 

This paper is organized as follows. We first present the TDDFPT formalism with SR US-PPs as implemented in two ways, the Sternheimer equations and the LL method, 
for periodic solids and finite momentum transfer (Section~\ref{sec:TDDFPT}). The LL method with  US-PPs is benchmarked against the NC-PPs implementation and the FP-LAPW method for bulk Au, and then we present our results, comparison with the experiments, and discuss the origin of the peaks in the EEL spectra of bulk Au in Section~\ref{sec:application}. 
Conclusions are drawn in section~\ref{sec:conclusion}.

\section{TDDFPT formalism with ultrasoft pseudopotentials}
\label{sec:TDDFPT} 

In the following, the formalism is detailed for insulators for the sake of simplicity and clarity and the discussion is limited to scalar relativistic PPs. 
The Reader is referred to ref.~\onlinecite{Dalcorso:2001} for the metallic case in US-PP DFPT.  
Hartree atomic units are used throughout the paper. 
For operators we will use a hat on top ``$\hat{\phantom{O}}$'' of the symbol.

\subsection{TDDFT equations in the US-PP scheme}\label{sec:uspp_sr}

Both EELS and IXS cross-sections are 
proportional to $S({\bf q},\omega)$, the dynamical structure factor per unit volume
of the solid, where ${\bf q}$ is the transferred momentum and $\omega$ is the energy
loss. $S({\bf q},\omega)$ is proportional to the imaginary part of the
charge-density susceptibility $\chi({\bf q}, {\bf q}, \omega)$:
\begin{equation}
S({\bf q},\omega) = - {1\over \pi} {\rm Im} \, \chi({\bf q}, {\bf q}, \omega).
\end{equation}
The charge-density susceptibility $\chi(\mathbf{r}, \mathbf{r}', \omega)$ of a system 
relates to the charge density induced by an external perturbing
potential~\cite{Botti:2007}:
\begin{equation}\label{eq:chi_density}
 n'(\mathbf{r},\omega) = \int d^3r' \chi (\mathbf{r},\mathbf{r}',\omega) V^{\prime}_{\mathrm{ext}}(\mathbf{r}',\omega) .
\end{equation} 
Therefore, when the external perturbation is an electron (plane wave) with a fixed momentum $\mathbf{q}$ and the external perturbing potential is $V^{\prime}_{\mathrm{ext}}(\mathbf{r}',\omega) = e^{i {\bf q}\cdot {\bf r}'}$ we see from Eq.~\eqref{eq:chi_density} that the charge-density response at frequency $\omega$ reads: 
\begin{equation} \label{eq:ncpp_chi}
n'(\mathbf{r},\omega)=  \chi (\mathbf{r},{\bf q},\omega) ,
\end{equation}
and the subsequent Fourier transform of $n'(\mathbf{r},\omega)$ at ${\bf q}$ is the
requested charge-density susceptibility 
$\chi({\bf q}, {\bf q}, \omega)$. 

In the time domain, the electronic charge density in Vanderbilt's US-PP scheme~\cite{Vanderbilt:1990} 
reads:
\begin{equation}\label{eq:uspp_density}
 n(\mathbf{r},t) = 2 \sum_{\mathbf{k},i} \Braket{\psi_{\mathbf{k},i}(t)|\hat K(\mathbf{r})|\psi_{\mathbf{k},i}(t)} ,
\end{equation}
where the index $\mathbf{k}$ runs over the points in the Brillouin zone (BZ), the index $i$ runs over the occupied Kohn-Sham (KS) wavefunctions, and the factor $2$ accounts for spin degeneracy. 
In Eq.~\eqref{eq:uspp_density}, $\hat K(\mathbf{r})$ is a nonlocal operator at every point in space $\mathbf{r}$ and in coordinate representation it would be $K(\mathbf{r},\mathbf{r}_1,\mathbf{r}_2)$~\cite{Dalcorso:2001}. In the NC-PP case, $K(\mathbf{r},\mathbf{r}_1,\mathbf{r}_2)$ is simply $\delta(\mathbf{r}-\mathbf{r}_1) \delta(\mathbf{r}-\mathbf{r}_2)$, with $\delta$ the Dirac distribution, and hence Eq.~\eqref{eq:uspp_density} reduces to $n(\mathbf{r},t) = 2 \sum_{\mathbf{k},i} |\psi_{\mathbf{k},i}(\mathbf{r}, t)|^2$. Instead, in the US-PP 
case,
it contains the so-called augmentation term due to the lift of the norm 
conservation constraint on the pseudo wavefunction:
\begin{equation}\label{eq:K_us_op}
 \hat K(\mathbf{r})=\ket{\mathbf{r}}\bra{\mathbf{r}}+\sum_{Imn} Q^{\gamma(I)}_{mn}(\mathbf{r}-\mathbf{R}_I) \ket{\beta^I_m}\bra{\beta^I_n},
\end{equation} 
where the index $I$ runs over atoms, $\gamma(I)$ is the type of atom $I$, $Q^{\gamma(I)}_{mn}(\mathbf{r} -\mathbf{R}_I)$ and $\langle {\bf r} |\beta^I_m\rangle \equiv \beta^{\gamma(I)}_m(\mathbf{r}-\mathbf{R}_I)$ are the augmentation functions and projector functions of atom $I$ centered at $\mathbf{R}_I$, respectively, and the indices $m$ and $n$ run over all the projectors of the atom $I$. $Q^{\gamma(I)}_{mn}(\mathbf{r})$ are calculated by pseudizing the difference 
${\varphi}^{\gamma(I) \, *}_m(\mathbf{r}){\varphi}^{\gamma(I)}_n(\mathbf{r}) - \phi^{\gamma(I) \, *}_m(\mathbf{r}) \phi^{\gamma(I)}_n(\mathbf{r})$ so that they 
are easily expanded in plane waves but 
conserve the multipole moments.
Here ${\varphi}^{\gamma(I)}_m(\mathbf{r})$ and $\phi^{\gamma(I)}_n(\mathbf{r})$
are the all-electron and pseudo partial waves, respectively~\cite{Vanderbilt:1990}.
The augmentation functions and the projector functions are localized in spheres 
about each atom $I$ and are generated together with the US-PP. 

In the US-PP scheme of TDDFPT, the TD KS equations read:~\cite{Qian:2006}
\begin{equation}\label{eq:tddfpt_ks_US}
i \hat{S} \, \frac{\partial |\psi_{\mathbf{k},i}(t)\rangle}{\partial t}=
\hat{H}_\mathrm{KS}(t) \, |\psi_{\mathbf{k},i}(t)\rangle ,
\end{equation}
where $\hat{S}$ is an overlap operator:
\begin{equation}\label{eq:S_op}
\hat{S}=1+\sum_{Imn} q^{\gamma(I)}_{mn}\ket{\beta^I_m}\bra{\beta^I_n}, 
\end{equation}
whose coefficients are defined as $q^{\gamma(I)}_{mn} = \int d^3r \, Q^{\gamma(I)}_{mn}(\mathbf{r})$.  In Eq.~\eqref{eq:tddfpt_ks_US}, $\hat{H}_{\mathrm{KS}}(t)$ is the 
TD KS Hamiltonian which reads:
\begin{equation}\label{eq:tddfpt_ham}
\hat{H}_{\mathrm{KS}}(t) = \hat{H}^\circ + \hat{V}'(t) .
\end{equation}
It is a nonlocal operator, where $\hat H^\circ$ is the Hamiltonian of the unperturbed 
system, and $\hat{V}'(t)$ is the TD linearized potential:
\begin{equation}\label{eq:response_potential}
\hat{V}'(t) = \int d^3r'
\left[ V'_{\mathrm{ext}}(\mathbf{r}',t) + 
V'_{\mathrm{Hxc}}(\mathbf{r}',t)\right] \hat K(\mathbf{r}') \, .
\end{equation}
Here, $V'_\mathrm{ext}(\mathbf{r},t)$ is the external TD perturbing potential, and $V'_{\mathrm{Hxc}}(\mathbf{r},t)$ is the linear-response TD Hartree and exchange-and-correlation (Hxc) potential. At variance with the NC-PPs case, in the US-PPs case
the operator $\hat{V}'(t)$ is nonlocal, because 
$\hat{K}(\mathbf{r})$ is nonlocal [see Eq.~\eqref{eq:K_us_op}]: $\hat{V}'(t)$ 
in the coordinate representation is $V'(\mathbf{r}_1, \mathbf{r}_2, t)$.
We consider a real external perturbation of the form\cite{Timrov:2013}:
\begin{eqnarray}
V'_{\mathrm{ext}}(\mathbf{r},t) & = & \int\limits_{-\infty}^{\infty} d\omega \,
V'_\mathrm{ext}(\mathbf{r},\omega) \, e^{-i\omega t}  \nonumber \\
&=&
\int\limits_{0}^{\infty} d\omega \, \left[ V'_\mathrm{ext}(\mathbf{r},\omega) \, e^{-i\omega t} 
+ c.c. \right] ,
\label{eq:EELS_Vext}
\end{eqnarray}
and compute the linearly induced charge density $n{'}(\mathbf{r},t)$. 
The TD KS wavefunctions can be developed to first order as:
\begin{equation}\label{eq:wfc_srnc}
| \psi_{\mathbf{k},i}(t)\rangle = e^{-i \varepsilon_{\mathbf{k},i} t} \, \left[ 
|\psi^\circ_{\mathbf{k},i}\rangle + |\psi'_{\mathbf{k},i}(t)\rangle\right] ,
\end{equation}
so that the TD charge density becomes $n(\mathbf{r},t)=n^{\circ}(\mathbf{r})+n{'}(\mathbf{r},t)$,
where $n^\circ(\mathbf{r})$ is the unperturbed charge density. 
Going from time to frequency domain by Fourier transforming all quantities, the Fourier transform of $n{'}(\mathbf{r},t)$ reads:
\begin{widetext}
\begin{eqnarray}\label{eq:response_cd}
n'(\mathbf{r},\omega) & = & 2 \sum_{\mathbf{k},i} \biggl\{ \langle \psi^{\circ}_{\mathbf{k},i}|\mathbf{r}\rangle\langle\mathbf{r}|\psi'_{\mathbf{k},i}(\omega)\rangle+
\langle \psi'_{\mathbf{k},i}(-\omega)|\mathbf{r}\rangle\langle \mathbf{r}|
\psi^\circ_{\mathbf{k},i}\rangle \biggr. \nonumber \\
& + & \biggl. \sum_{Imn} Q^{\gamma(I)}_{mn}(\mathbf{r}-\mathbf{R}_I) \Bigl[
 \braket{\psi^\circ_{\mathbf{k},i}|\beta^I_m} \braket{\beta^I_n|
\psi'_{\mathbf{k},i}(\omega)} + 
\braket{\psi'_{\mathbf{k},i}(-\omega)|\beta^I_m}
\braket{\beta^I_n|\psi^\circ_{\mathbf{k},i}}
 \Bigr] \biggr\} \,,
\end{eqnarray}
\end{widetext}
where $|\psi'_{\mathbf{k},i}(\omega)\rangle$ is the Fourier transform of  $|\psi'_{\mathbf{k},i}(t)\rangle$.
The formalism of the NC-PPs can be recovered by setting the augmentation terms to zero, \textit{i.e.} only the first row in the equation above will remain~\cite{Timrov:2013}.

In a periodic solid it is convenient to use the Bloch theorem 
by writing the KS wavefunctions as:
$\langle\mathbf{r}|\psi^\circ_{\mathbf{k},i}\rangle=e^{i\mathbf{k}\cdot\mathbf{r}} 
\langle \mathbf{r} |u^\circ_{\mathbf{k},i}\rangle$ ,
where $\langle \mathbf{r}|u^\circ_{\mathbf{k},i}\rangle$ is a lattice-periodic function. 
The total external perturbing potential $V'_{\mathrm{ext}}(\mathbf{r},\omega)$ can 
be written as a sum of the Fourier monochromatic $\mathbf{q}$ components, \textit{i.e} as:
\begin{equation}
V'_{\mathrm{ext}}(\mathbf{r},\omega) = \sum_\mathbf{q} e^{i\mathbf{q}\cdot\mathbf{r}} \, v'_{\mathrm{ext},\mathbf{q}}(\mathbf{r},\omega) ,
\label{eq:Vext_q_decomposition}
\end{equation}
where $v'_{\mathrm{ext},\mathbf{q}}(\mathbf{r},\omega)$ is the lattice-periodic part of the perturbation. 
In EELS, for a beam of incoming electrons, each of which undergoing  a 
certain momentum transfer $\mathbf{q}$, the perturbation is  
$v'_{\mathrm{ext},\mathbf{q}}(\mathbf{r},\omega)=1$ for a given ${\bf q}$, 
and zero for all the others. In this case, the response KS wavefunctions 
can be written as: 
\begin{equation}
\langle\mathbf{r}|\psi'_{\mathbf{k},i}(\omega)\rangle = \sum_{\mathbf{q}} e^{i(\mathbf{k+q})\cdot \mathbf{r}} \,
\langle\mathbf{r}|u'_{\mathbf{k}+\mathbf{q},i}(\omega)\rangle \,.
\end{equation}
The response charge density and response HXC potential  
can be decomposed in the same way:
\begin{equation}
n'(\mathbf{r},\omega) = \sum_{\mathbf{q}} e^{i\mathbf{q}\cdot\mathbf{r}} \, {n}'_\mathbf{q}(\mathbf{r},\omega) ,
\label{eq:EELS_charge-dens-response_5}
\end{equation}
where ${n}'_\mathbf{q}(\mathbf{r},\omega)$ is the lattice-periodic part. 
After introducing the identity $1 = \hat P_v + \hat P_c$ in Eq.~\eqref{eq:response_cd}, with $\hat P_c$ (resp. $\hat P_v$) the projectors onto
the conduction (resp.~valence) states, 
it can be shown that ${n}'_{\mathbf{q}}({\bf r},\omega)$  reads: 
\begin{widetext}
\begin{eqnarray}
n'_\mathbf{q}(\mathbf{r},\omega) & = & 2 \sum_{\mathbf{k},i} \biggl\{
\langle u^{\circ}_{\mathbf{k},i}|\mathbf{r} \rangle \langle\mathbf{r}|\hat P_c^{\mathbf{k+q}}u'_{\mathbf{k+q},i}(\omega)\rangle+ \langle u^{\circ}_{\mathbf{k},i}|\mathbf{r} \rangle \langle\mathbf{r}| \hat P_c^{\mathbf{k+q}} u^{\prime \, *}_{\mathbf{-k-q},i}(-\omega)\rangle \biggr. \nonumber \\
& + & \biggl. \sum_{smn} \tilde{Q}^{\gamma(s), \mathbf{q}}_{mn}(\mathbf{r} - {\boldsymbol \tau}_s) 
\langle \psi_{\mathbf{k},i}|\beta_{m}^{s} \rangle
\Bigl[ A^{s,\mathbf{k+q},i}_{n}(\omega) + B^{s,\mathbf{-k-q},i}_{n}(-\omega) \Bigr]
\biggr\},
\label{induced_n}
\end{eqnarray}
where
\begin{equation}
A^{s,\mathbf{k+q},i}_{n}(\omega) = \int d^3r \, \beta^{s*}_n (\mathbf{r} - {\boldsymbol \tau}_s) \, e^{i\mathbf{(k+q)}\cdot \mathbf{r}} \, \langle\mathbf{r}|\hat P_c^{\mathbf{k+q}}u'_{\mathbf{k+q},i}(\omega)\rangle,
\end{equation}
\begin{equation}
B^{s,\mathbf{-k-q},i}_{n}(-\omega)= \int d^3r \, \beta^{s*}_n (\mathbf{r} - {\boldsymbol \tau}_s) \, e^{i\mathbf{(k+q)}\cdot\mathbf{r}} \, \langle\mathbf{r}|\hat P_c^{\mathbf{k+q}}u^{\prime \, *}_{\mathbf{-k-q},i}(-\omega)\rangle,
\end{equation}
\end{widetext}
and
\begin{equation}
\tilde{Q}^{\gamma(s), \mathbf{q}}_{mn}(\mathbf{r} - {\boldsymbol \tau}_s) =
e^{-i\mathbf{q}\cdot\mathbf{r}} \sum_l e^{i\mathbf{q}\cdot\mathbf{R}_l} Q^{\gamma(s)}_{mn}(\mathbf{r}-\mathbf{R}_l - {\boldsymbol \tau}_s) . 
\label{eq:Q_periodic}
\end{equation}
In Eqs.~\eqref{induced_n} -- \eqref{eq:Q_periodic}, and in the following, we use the fact that atomic positions in a periodic solid can be indicated as $\mathbf{R}_I=
\mathbf{R}_l + {\boldsymbol \tau}_s$, where $\mathbf{R}_l$ is a Bravais lattice 
vector 
and ${\boldsymbol \tau}_s$ is the position of the atom in one unit cell 
($I=\{l,s\}$). 
The projector onto the conduction manifold  $\hat{P}_c^{\mathbf{k+q}}$ is defined as~\cite{Baroni:2001,Dalcorso:2001}:
\begin{eqnarray}
\hat{P}_c^{\mathbf{k+q}}= 1 - \sum_{j}^\mathrm{occ} |u^\circ_{\mathbf{k+q},j}\rangle
\langle u^{\circ}_{\mathbf{k+q},j}| \, \hat{S}_\mathbf{k+q} \,,
\end{eqnarray}
where the sum over $j$ runs over the occupied states. The overlap operator
$\hat{S}_\mathbf{k+q}$ in the coordinate representation is defined as:
\begin{equation}
\langle \mathbf{r} |\hat{S}_\mathbf{k+q}|\mathbf{r}'\rangle = 
e^{-i (\mathbf{k+q})\cdot\mathbf{r}} \, \langle\mathbf{r}| \hat{S} |\mathbf{r}'\rangle
e^{i (\mathbf{k+q})\cdot\mathbf{r}'}. 
\end{equation}
We have used the same notations as in Eq.~(34) of ref.~\onlinecite{Baroni:2001}. 

\subsection{The Sternheimer equations in the US-PP scheme}

The responses of the wavefunctions, 
$|\hat P_c^{\mathbf{k+q}}u'_{\mathbf{k+q},i}(\omega)\rangle$ and 
$|\hat P_c^{\mathbf{k+q}}u'^*_{\mathbf{-k-q},i}(-\omega)\rangle$, that appears in 
Eq.~\eqref{induced_n} can be obtained, within TDDFPT, 
by solving the Sternheimer equations. In refs.~\onlinecite{Timrov:2013,Timrov:2013b,Timrov:2013d,Timrov:2015} these
equations have been derived with NC-PPs.  
By inserting Eq.~\eqref{eq:wfc_srnc} in Eq.~\eqref{eq:tddfpt_ks_US}, using the 
Bloch theorem, and making a Fourier transformation from the time domain to the frequency domain, we obtain the first Sternheimer equation for the lattice-periodic part of the response KS wavefunctions in the US-PP scheme: 
\begin{multline}\label{eq:stern_us_1_kq}
\left[ \hat H^{\circ}_{\mathbf{k+q}}-(\varepsilon_{\mathbf{k},i}+\omega)
\hat{S}_{\mathbf{k+q}}\right] |\hat P_c^{\mathbf{k+q}}u'_{\mathbf{k+q},i}(\omega)
\rangle = \\
-\hat P_c^{\dagger \, \mathbf{k+q}} \,
\hat{v}'_\mathbf{q}(\omega) |u^\circ_{\mathbf{k},i}\rangle.  
\end{multline}
The Hermitian conjugation in the operator $\hat P_c^{\dagger \, \mathbf{k+q}}$ comes from the presence of the overlap matrix with US-PPs and has no equivalence in the NC-PP case. The operator $\hat{H}^\circ_\mathbf{k+q}$ is defined as \cite{Baroni:2001}: 
\begin{equation}
\langle \mathbf{r}|\hat{H}^\circ_\mathbf{k+q}|\mathbf{r'}\rangle = 
e^{-i (\mathbf{k+q})\cdot\mathbf{r}} \langle \mathbf{r}| \hat{H}^\circ | 
\mathbf{r'}\rangle e^{i (\mathbf{k+q})\cdot\mathbf{r'}}.
\end{equation}
Considering the complex conjugate of Eq.~\eqref{eq:stern_us_1_kq} at $-\mathbf{k}$ for a perturbation with $-\mathbf{q}$ and $-\omega$, and by using the time-reversal symmetry, we obtain the second Sternheimer equation:
\begin{multline}\label{eq:stern_us_2_kq}
\left[ \hat{H}^{\circ}_{\mathbf{k+q}}-(\varepsilon_{\mathbf{k},i}-\omega)\hat{S}_{\mathbf{k+q}}\right] |\hat P_c^{\mathbf{k+q}} u'^*_{\mathbf{-k-q}}(-\omega)\rangle = \\
-\hat P_c^{\dagger \, \mathbf{k+q}} \,
\hat{v}'_\mathbf{q}(\omega) |u^\circ_{\mathbf{k},i}\rangle .
\end{multline}
Here, due to time-reversal symmetry,
 $v^{\prime \, *}_{\mathrm{Hxc},-\mathbf{q}}(\mathbf{r},-\omega) =
v'_{\mathrm{Hxc},\mathbf{q}}(\mathbf{r},\omega)$, and we also used the fact that
$\hat K(\mathbf{r})$ is a real operator. 
Except for the presence of the overlap matrix $\hat{S}_{\mathbf{k+q}}$, 
Eqs.~\eqref{eq:stern_us_1_kq}~and~\eqref{eq:stern_us_2_kq} 
are formally similar to the NC-PP case
presented in refs.~\onlinecite{Timrov:2013,Timrov:2013b,Timrov:2013d,Timrov:2015}. 
Note, however, that in the US-PP case, the frequency-dependent potential on the right-hand side of Eqs.~\eqref{eq:stern_us_1_kq}~and~\eqref{eq:stern_us_2_kq}
is a nonlocal operator and has a more complex form due to the augmentation 
terms. Consequently, the object $\hat{v}'_\mathbf{q}(\omega)|
u^\circ_{\mathbf{k},i}\rangle$ must be interpreted as a shorthand notation
for the lattice-periodic part of $\hat{V}'(\omega)|\psi^\circ_{\mathbf{k},i}\rangle$: 
\begin{eqnarray}\label{eq:dvsfc_us}
& & \langle \mathbf{r}|\hat{v}'_\mathbf{q}(\omega)|
u^\circ_{\mathbf{k},i}\rangle \equiv e^{-i\mathbf{(k+q)}\cdot\mathbf{r}}
\nonumber \\
& & \times \, \langle \mathbf{r}|
\int d^3r' \, e^{i\mathbf{q}\cdot\mathbf{r}'} (1 + \hat{v}'_{\mathrm{Hxc},\mathbf{q}}(\mathbf{r}',\omega))
\hat K(\mathbf{r}') |\psi^\circ_{\mathbf{k},i}\rangle .
\end{eqnarray}
By inserting the expression of $\hat K(\mathbf{r}')$ [see Eq.~\eqref{eq:K_us_op}] in Eq.~\eqref{eq:dvsfc_us}, we can rewrite Eqs. (22) and (24) as:
\begin{widetext}
\begin{eqnarray}\label{stern1_w_us_final}
\Big[\hat H^\circ_{\mathbf{k+q}}-(\varepsilon_{\mathbf{k},i} + \omega)\hat{S}_{\mathbf{k+q}}\Big]
|\hat P_c^{\mathbf{k+q}}u'_{\mathbf{k+q},i}(\omega)\rangle
= & - & \hat P_c^{\dagger \, \mathbf{k+q}} \Big[ |u^\circ_{\mathbf{k},i}\rangle+
\hat{v}'_{\mathrm{Hxc},\mathbf{q}}(\omega) 
|u^\circ_{\mathbf{k},i}\rangle \nonumber \\
& + & \sum_{smn} \left(^0I^{s,\mathbf{q}}_{mn} +
^3I^{s,\mathbf{q}}_{mn}(\omega) \right)
|\beta_m^{s, \mathbf{k+q}} \rangle  \langle \beta^{s}_n|\psi^\circ_{\mathbf{k},i}\rangle
\Big],
\end{eqnarray}
\begin{eqnarray}\label{stern2_w_us_final}
\Big[\hat{H}^\circ_{\mathbf{k+q}}-(\varepsilon_{\mathbf{k},i}-\omega)\hat{S}_{\mathbf{k+q}}\Big]
|\hat P_c^{\mathbf{k+q}} u'^*_{\mathbf{-k-q},i}(-\omega)\rangle
= & - & \hat P_c^{\dagger \, \mathbf{k+q}} \Big[ |u^\circ_{\mathbf{k},i}\rangle+ 
\hat{v}'_{\mathrm{Hxc},\mathbf{q}}(\omega) 
|u^\circ_{\mathbf{k},i}\rangle \nonumber \\
& + & \sum_{smn} \left( ^0I^{s,\mathbf{q}}_{mn}
+ ^3I^{s,\mathbf{q}}_{mn}(\omega) \right)
|\beta_m^{s, \mathbf{k+q}} \rangle 
\langle \beta^s_n|\psi^\circ_{\mathbf{k},i}\rangle \Big] ,
\end{eqnarray}
\end{widetext}
where we defined $^0I^{s,\mathbf{q}}_{mn}$ as: 

\begin{equation}
 ^0I^{s,\mathbf{q}}_{mn} = \int d^3r \, e^{i\mathbf{q}\cdot\mathbf{r}} \, 
 Q^{\gamma(s)}_{mn}(\mathbf{r} - {\boldsymbol \tau}_s) ,
\end{equation}
and $^3I^{s,\mathbf{q}}_{mn}(\omega)$ as:
\begin{equation}
 ^3I^{s,\mathbf{q}}_{mn}(\omega)=\int d^3r \, e^{i\mathbf{q}\cdot\mathbf{r}} \,
v'_{\mathrm{Hxc},\mathbf{q}}(\mathbf{r},\omega) Q^{\gamma(s)}_{mn}(\mathbf{r} - {\boldsymbol \tau}_s) .
\end{equation}
We have used the notations $^0I^{s,\mathbf{q}}_{mn}$ and $^3I^{s,\mathbf{q}}_{mn}(\omega)$ to be consistent with notations in ref.~\onlinecite{Dalcorso:2001}. 
Moreover, we have defined [cf. with Eq.~\eqref{eq:Q_periodic}]:
\begin{equation}
\langle \mathbf{r}|\beta_m^{s, \mathbf{k+q}}\rangle =  e^{-i(\mathbf{k+q})\cdot\mathbf{r}} 
\sum_{l} e^{i(\mathbf{k+q})\cdot\mathbf{R}_l} \beta^{\gamma(s)}_m(\mathbf{r} - \mathbf{R}_l -
{\boldsymbol \tau}_s). 
\label{eq:beta_kq}
\end{equation}
Note that $\langle\beta^s_n|\psi^o_{{\bf k },i}\rangle=
{1\over N} \langle\beta^{s,{\bf k}}_n|u^o_{{\bf k },i}\rangle$, where $N$ is the number of cells of the system,  so that
Eqs.~\eqref{stern1_w_us_final}~and~\eqref{stern2_w_us_final} are written in terms of lattice-periodic functions.
The linear system of Eqs.~\eqref{stern1_w_us_final} and \eqref{stern2_w_us_final} can be solved self-consistently at each frequency~$\omega$. 
The Fourier transform of the lattice-periodic part of the self-consistent response charge-density
calculated at $\mathbf{G=0}$ yields  the required susceptibility, 

\begin{equation}
n'(\mathbf{q+G},\omega) = \sum_{\mathbf{G}'} \chi (\mathbf{q+G},\mathbf{q+G'},\omega) \, V'_{\mathrm{ext}}(\mathbf{q+G'},\omega), 
\end{equation}
and in order to obtain the macroscopic dielectric function we only need terms for which $\mathbf{G}=\mathbf{G}'=0$. Finally, the electronic susceptibility is given by Eq.~\eqref{eq:ncpp_chi}.

 \subsection{The quantum Liouville equation in the US-PP scheme}\label{sec:q_liouv_uspp}
 
 An alternative approach to the self-consistent solution of the Sternheimer equations 
at each frequency has been developed using the quantum Liouville equation, the 
so-called LL method~\cite{Walker:2006,Walker:2007,Malcioglu:2011,Rocca:2008,Timrov:2013,Timrov:2013b,Timrov:2013d,Timrov:2015} that allows a significant reduction of the 
required computational resources. 
In this approach one can work in the standard batch representation (SBR),
where the response KS wavefunctions are rotated and represented as batches~$q_\mathbf{q} = \{ |q_{\mathbf{k+q},i}\rangle \}$ and $p_\mathbf{q} = \{ |p_{\mathbf{k+q},i}\rangle \}$, where
\begin{equation}
|q_{\mathbf{k+q},i}\rangle = \frac{1}{2} \, [ |
\hat P_c^{\mathbf{k+q}}{u}'_{\mathbf{k+q},i}(\omega)\rangle +|\hat P_c^{\mathbf{k+q}} {u}'^*_{\mathbf{-k-q},i}(-\omega)\rangle ] ,
\label{eq:EELS_Batch_q}
\end{equation}
\begin{equation}
|p_{\mathbf{k+q},i}\rangle= \frac{1}{2} \, [ |
\hat P_c^{\mathbf{k+q}}{u}'_{\mathbf{k+q},i}(\omega)\rangle -|\hat P_c^{\mathbf{k+q}} {u}'^*_{\mathbf{-k-q},i}(-\omega)\rangle ].
\label{eq:EELS_Batch_p}
\end{equation}
In the SBR the linearized quantum Liouville equation in the US-PP scheme can
be derived by writing the Sternheimer equations~\eqref{stern1_w_us_final} and \eqref{stern2_w_us_final} in terms of $|q_{\mathbf{k+q},i}\rangle$ and $|p_{\mathbf{k+q},i}\rangle$. 
Multiplying both equations by $\hat{S}^{-1}_\mathbf{k+q}$ we obtain (see appendix~\ref{app:Sm1}): 

\begin{equation}
\left(
\begin{array}{cc}
\omega & -\hat{\mathcal{D}}_\mathbf{q} \\
-\hat{\mathcal{D}}_\mathbf{q}-\hat{\mathcal{K}}_\mathbf{q} & \omega
\end{array}
\right)
\left(
\begin{array}{c}
q_\mathbf{q} \\
p_\mathbf{q}
\end{array}
\right) = 
\left(
\begin{array}{c}
0 \\
y_\mathbf{q}
\end{array}
\right) ,
\label{eq:EELS_Liouvillian_eq_SBR_us}
\end{equation}
\\
where the actions of the operators $\hat{\mathcal{D}}_\mathbf{q}$ and $\hat{\mathcal{K}}_\mathbf{q}$ on the batches are defined as:
\begin{equation}
\hat{\mathcal{D}}_\mathbf{q} \cdot q_\mathbf{q}  = \left \{ (\hat{S}^{-1}_\mathbf{k+q}\hat{H}^\circ_\mathbf{k+q}  - \varepsilon_{\mathbf{k},i}) \, |q_{\mathbf{k+q},i}\rangle
\right \} ,
\label{eq:EELS_D_super-op_us}
\end{equation}
and
\begin{eqnarray}\label{eq:EELS_K_super-op_us}
\hat{\mathcal{K}}_\mathbf{q}\cdot q_\mathbf{q} & = & \biggl\{
\hat{S}^{-1}_\mathbf{k+q}\hat{P}_c^{\dagger \, \mathbf{k+q}} \Bigl[
\hat{v}'_{\mathrm{Hxc},\mathbf{q}}(\omega) 
|u^\circ_{\mathbf{k},i}\rangle \Bigr. \biggr. \nonumber \\
& & \biggl. \Bigl. +  \sum_{smn}
\ ^3I^{s,\mathbf{q}}_{mn}(\omega)\ 
|\beta_m^{s, \mathbf{k+q}} \rangle \langle \beta^{s}_m|
\psi^\circ_{\mathbf{k},i}\rangle \Bigr] \biggr\}.
\end{eqnarray}
On the right-hand side of Eq.~\eqref{eq:EELS_Liouvillian_eq_SBR_us}, $y_\mathbf{q}$ is the perturbation term which reads:
\begin{eqnarray}\label{eq:EELS_K_super-op_rhs}
y_\mathbf{q} & = & \biggl\{ \hat{S}^{-1}_\mathbf{k+q} \hat{P}_c^{\dagger \mathbf{k+q}} 
\Bigl[ |u^\circ_{\mathbf{k},i} \rangle \Bigr. \biggr. \nonumber \\
& & \biggl. \Bigl. + \sum_{smn} \ ^0I^{s,\mathbf{q}}_{mn}\ |\beta_m^{s, \mathbf{k+q}} \rangle  \langle 
\beta^s_m|\psi^\circ_{\mathbf{k},i}\rangle \Bigr] \biggr\} .
\end{eqnarray}
If the exchange-correlation kernel is adiabatic, 
$\hat{v}'_{\mathrm{Hxc},\mathbf{q}}(\omega)$ and 
$^3I^{s,\mathbf{q}}_{mn}(\omega)$ depend on $\omega$ only implicitly through the
$q_\mathbf{q}$ batch. 
Actually, Eq.~\eqref{induced_n} can be rewritten in terms of $q_{\mathbf{q}}$ as
\begin{eqnarray}
n'_\mathbf{q}(\mathbf{r},\omega)&=&4\sum_{\mathbf{k},i}\Big\{
\langle u^{\circ}_{\mathbf{k},i}|\mathbf{r} \rangle \langle\mathbf{r}|q_{\mathbf{k+q},i}
\rangle \nonumber \\
 & & \hspace{-1.5cm} + \sum_{smn} \tilde{Q}^{\gamma(s),\mathbf{q}}_{mn}(\mathbf{r} - {\boldsymbol \tau}_s) 
\langle \psi^{\circ}_{\mathbf{k},i}|\beta^{s}_m \rangle  
C^{s,\mathbf{k+q},i}_{n}(\omega)
\Big\},
\label{induced_n1}
\end{eqnarray}
where
\begin{equation}
C^{s,\mathbf{k+q},i}_{n}(\omega)= \int d^3r \, \beta^{s*}_n (\mathbf{r} - {\boldsymbol \tau}_s) \, e^{i(\mathbf{k+q})\cdot\mathbf{r}} \, \langle\mathbf{r}|q_{\mathbf{k+q},i}\rangle. 
\end{equation}
In this case, the susceptibility $\chi(\mathbf{q},\mathbf{q},\omega)$ is given by the following expression~\cite{Timrov:2013b,Timrov:2013d}:
\begin{equation}
\chi(\mathbf{q},\mathbf{q},\omega) = \left\langle (y_\mathbf{q}, 0) \Big| \left(
\begin{array}{cc}
\omega & -\hat{\mathcal{D}}_\mathbf{q} \\
-\hat{\mathcal{D}}_\mathbf{q}-\hat{\mathcal{K}}_\mathbf{q} & \omega
\end{array}
\right)^{-1}
\left(
\begin{array}{c}
0 \\
y_\mathbf{q}
\end{array}
\right) \right\rangle ,
\end{equation}
and Eq.~\eqref{eq:EELS_Liouvillian_eq_SBR_us} can be solved using the Lanczos recursive algorithm~\cite{Walker:2006,Walker:2007,Malcioglu:2011,Rocca:2008,Timrov:2013,Timrov:2013b,Timrov:2013d,Timrov:2015} identical to that used for the NC-PP case for any desired range and number 
of frequencies at the same computational cost, irrespective of the number of frequencies.
This is important when semicore states need to be included in the calculation  and treated 
as valence states, in order to  extend the frequency range on which the EEL spectrum is computed. 
We point out that in gold changes brought in by the introduction of semicore states in the valence show up above 15~eV (see appendix~\ref{app:no_semicore}). 

We also note that the multiplication by $\hat{S}^{-1}_\mathbf{k+q}$ is crucial to obtain an
expression of the Liouvillian that has a form similar to the NC-PP case
and can be solved by the Lanczos method~\cite{Gebauer:2013}. 
In other words, we use $\hat{S}^{-1}_\mathbf{k+q}$ in order to have equations in which
the frequency $\omega$ enters as a parameter, such that we can tridiagonalize
the resulting Liouvillian (defined by Eqs.~\eqref{eq:EELS_D_super-op_us}~and~\eqref{eq:EELS_K_super-op_us}) independently
on the value of the frequency. Ultimately, in the postprocessing step, the
tridiagonal matrix is used to solve linear systems for various values
of frequency $\omega$ at a negligible computational cost.

\section{Applications: EEL spectra of bulk Au}
\label{sec:application} 

In the present work we present results obtained using the LL method and US-PPs. 
The EEL spectra of bulk Au at vanishing {\bf q} are presented, as well as the peak dispersion, 
and the origin of the peaks is discussed. 

\subsection{Implementations} 

The LL approach to EELS~\cite{Timrov:2013,Timrov:2013b,Timrov:2013d} has been implemented in
the \texttt{turboEELS} code~\cite{Timrov:2015}, which is being distributed with the \textsc{Quantum
ESPRESSO} suite~\cite{Giannozzi:2017}. 
The ultrasoft capabilities introduced in this paper are available as the release 6.3 of \textsc{Quantum
ESPRESSO} (stable version of \texttt{turboEELS} with US-PPs). The Sternheimer
approach to TDDFPT has been implemented in the private branch of the
Quantum ESPRESSO project contained in the \texttt{Thermo{\_}pw} code~\cite{Dalcorso:2019}.
Sternheimer functionalities, for both norm-conserving and ultrasoft pseudopotentials, will be made available with the next version of the official distribution of \textsc{Quantum ESPRESSO}~\cite{Giannozzi:2017}. 

In a previous work~\cite{Motornyi:2018,Motornyi:2018b}, we compared the  results and performance 
 of the LL and Sternheimer approaches using NC-PPs. We showed that the two approaches yield the same results and 
demonstrated that calculations performed using the LL method require a CPU time smaller than for calculations performed using
 Sternheimer's approach when a given frequency range needs to be considered~\cite{Motornyi:2018,Motornyi:2018b}.  
In the present work, we have verified that also in the US-PP case both codes and both approaches give the same results. Performance turned out to be in 
favor of the LL method and the gain comparable to the case of NC-PPs.  
We report results at vanishing {\bf q}, and point out that the method is well suited for finite {\bf q} by showing the peak dispersion for gold.

\subsection{Computational details}
\label{sec:comput_details}

The results reported in the present work were obtained in the scalar-relativistic approximation, 
within the local-density approximation (LDA) for the exchange-correlation energy, and using US-PPs for Au with 11 ($5d^{10} 6s^1$) or 19 ($5s^2 5p^6 5d^{10} 6s^1$)  
electrons included in the valence region~\cite{dMotornyi:Note:2019d}. Two projectors have been used for each of the $s$, $p$ and $d$ channels~\cite{Dalcorso:2014}.   
For the 19 electron PP, reference energies for the $s$ angular momentum consisted of the $5s$ and $6s$ energy levels (resp. $5p$ and $6p$ energies for the $p$ angular momentum)~\cite{Dalcorso:2014}.We have also used the  generalized gradient approximation (GGA) with the PBE functional~\cite{Perdew:1996,Perdew:1997} in order to have a close comparison with the PBE-based calculations of ref.~\onlinecite{Alkauskas:2013}. 

The experimental lattice parameter at room temperature, $a_0=4.08$~\AA,  was used~\cite{Kittel:1972}. A kinetic energy cutoff
of $20$~Ry 
was used for the plane-wave expansion of the pseudowave functions for 
the US-PP with $11$ electrons ($160$~Ry for the charge density) and
$55$~Ry was used for the plane-wave expansion of the pseudowave functions for the NC-PP and US-PP with $19$ electrons ($220$~Ry for the charge density). 
Note that the unusually
small kinetic energy cut-off for the US-PP of Au with $11$ electrons is sufficient
to converge the reported spectrum~\cite{bMotornyi:Note:2019b}, but might be
too low for the computation of other physical properties such as phonon frequencies or elastic constants.
A $\mathbf{k}$ point mesh containing $32\times 32 \times 32$  special 
Monkhorst-Pack points was used for integrations  
over the BZ. The Methfessel-Paxton smearing  
was set to 2~mRy.  
 
Spectra for a very small value of the transferred momentum have been obtained at $|\mathbf{q}|\approx 0.03$~\AA$^{-1}$ in the $\Gamma-X$ direction of the BZ. 
The dispersion has been computed along this direction. 
In order to obtain a converged EEL spectrum, $6\,000$ Lanczos iterations have been computed and 
the Lanczos coefficients have been extrapolated to $30\,000$ iterations with the constant extrapolation method~\cite{Timrov:2015}. 
The frequencies used to compute the EEL spectrum of bulk Au have an imaginary part of $10$~mRy  that yields a Lorentzian broadening of the peaks.  

To substantiate our analysis, calculations in the random phase approximation (RPA) neglecting local field effects have been performed with the \textsc{SIMPLE} code at \textbf{q}=\textbf{0} exactly~\cite{Prandini:2019}, with the 19-electron norm-conserving pseudopotential of ref.~\onlinecite{Schlipf:2015}. We detail the calculations made with the \textsc{SIMPLE} code. 
Both the full response and the contribution of only interband transitions (ITs) have been obtained. 
The bulk dielectric function $\epsilon_B$  at \textbf{q}=\textbf{0} 
has been obtained by summing vertical (\textbf{q}=\textbf{0}) transitions, \textit{i.e.}
summing matrix elements of the dipolar operator between an occupied state and an empty one~\cite{Prandini:2019}. 
The intraband contribution has been added separately~\cite{Prandini:2019}.  
The loss function has then been obtained as: 
\begin{equation}\label{eq:loss}
-\mathrm{Im}(\epsilon_B^{-1}) = \frac{\mathrm{Im}(\epsilon_B)}{\mathrm{Re}(\epsilon_B)^2 + \mathrm{Im}(\epsilon_B)^2} ,
\end{equation}
where $\epsilon_B$ is the dielectric function of the bulk material without local fields effects, either containing both interband and intraband contributions, 
or being restricted to ITs. 
 
Finally, in order to identify plasmon peak positions, we present the loss function of two toy models computed with LL 
using modified pseudopotentials: a just 1-electron PP where $5d$ electrons are frozen in the core and a 10-electron PP from which 
the $6s$ electron was removed. These pseudopotentials were NC-PPs and designed by us by reducing the number of electrons 
in the generation of the 11-electron NC-PP of ref.~\onlinecite{Motornyi:Note:2017b}.   
 
The data used to produce the results of this work is available on the Materials Cloud Archive~\cite{fMotornyi:Note:2019f}.

\subsection{Benchmark of the Liouville-Lanczos implementation with US-PPs}
\label{sec:appendix_comp_NC_USPP}

In the present section we present a validation of our implementation of the LL approach with US-PPs. The EEL spectrum obtained in TDDFPT-GGA with the 19-electron US-PP developed in the present work (solid blue line), 
closely agree with our results obtained with the 19-electron NC-PP of ref.~\onlinecite{Schlipf:2015} (dashed black line), and compare well with a previous work~\cite{Alkauskas:2013}, 
where the loss function was obtained through the solution of the Dyson-like equation~\cite{Onida:2002} with an all-electron full-potential linearized augmented plane-wave (FP-LAPW) method (red dashed line).

\begin{figure}[h!]
  \centering
  \includegraphics[width=2.5in, angle=-90]{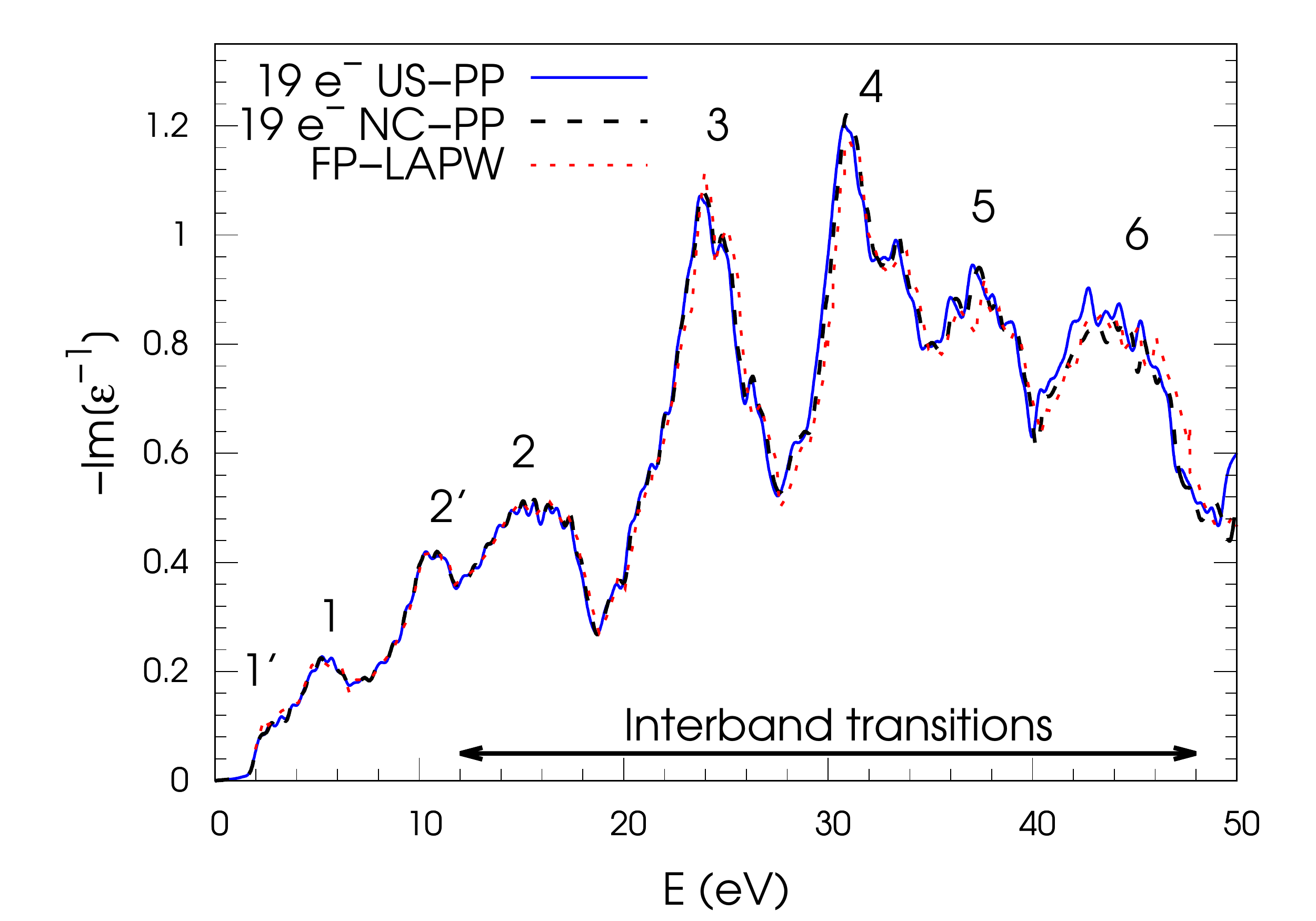}
  \caption{\label{fig:bulk_NC-PP} EEL spectrum of bulk Au computed with 19~electrons in the valence region, with ultrasoft and norm-conserving pseudopotentials, for $|\mathbf{q}|\approx 0.03$~\AA$^{-1}$ in the~[100] direction. Previous FP-LAPW data~\cite{Alkauskas:2013} is reported for $\mathbf{q}=0$. 
Peaks are labeled according to ref.~\onlinecite{Alkauskas:2013} and we have added the label 1$'$ and~2$'$. The horizontal arrow marks the frequency domain of pure interband transitions. Computations performed in TDDFPT-GGA. }
\end{figure}

The overall remarkable agreement between the EEL spectrum obtained using US-PP with another calculation (fig.~\ref{fig:bulk_NC-PP}) on a wide energy range confirms the correctness of the implementation of the LL approach with US-PPs. It is important to note though that special care must be given to the selection of US-PPs with high accuracy and transferability if one is interested in computing EEL spectra on a wide energy range,  as done in the present study.

The implementation of the LL approach with US-PPs can allow to achieve better performance with respect to NC-PPs thanks to the reduction of the cutoff and hence reduction of the CPU time. This is though element dependent. Unfortunately in Au this cannot be seen because the same value of the cutoff is needed for both types of PPs (see Sec.~\ref{sec:comput_details}), however calculations of plasmons in other elements, for which US-PPs are in bigger contrast with NC-PPs due to the hardness of the latter, can benefit more from our implementation of the LL approach with US-PPs. More specifically, the lower the cutoff value, the smaller the number of Lanczos iterations necessary to reach convergence of the EEL spectrum: this is a property of the LL approach~\cite{Rocca:2008}. Moreover, not only the number of Lanczos iterations can be decreased, but also the cost of each iteration will be reduced, which overall allows us to lower the CPU time requirements despite the fact that some extra-computations are needed because of the presence of the US-PP-related terms. As a matter of fact, the LL approach with US-PPs is widely used for the optical absorption spectroscopy of molecules~\cite{Walker:2007, Rocca:2008}.

\subsection{EEL spectra of bulk Au}
\label{subsec:EEL_bulk}

\begin{figure}[t]
    \centering
  \includegraphics[width=2.5in, angle=-90]{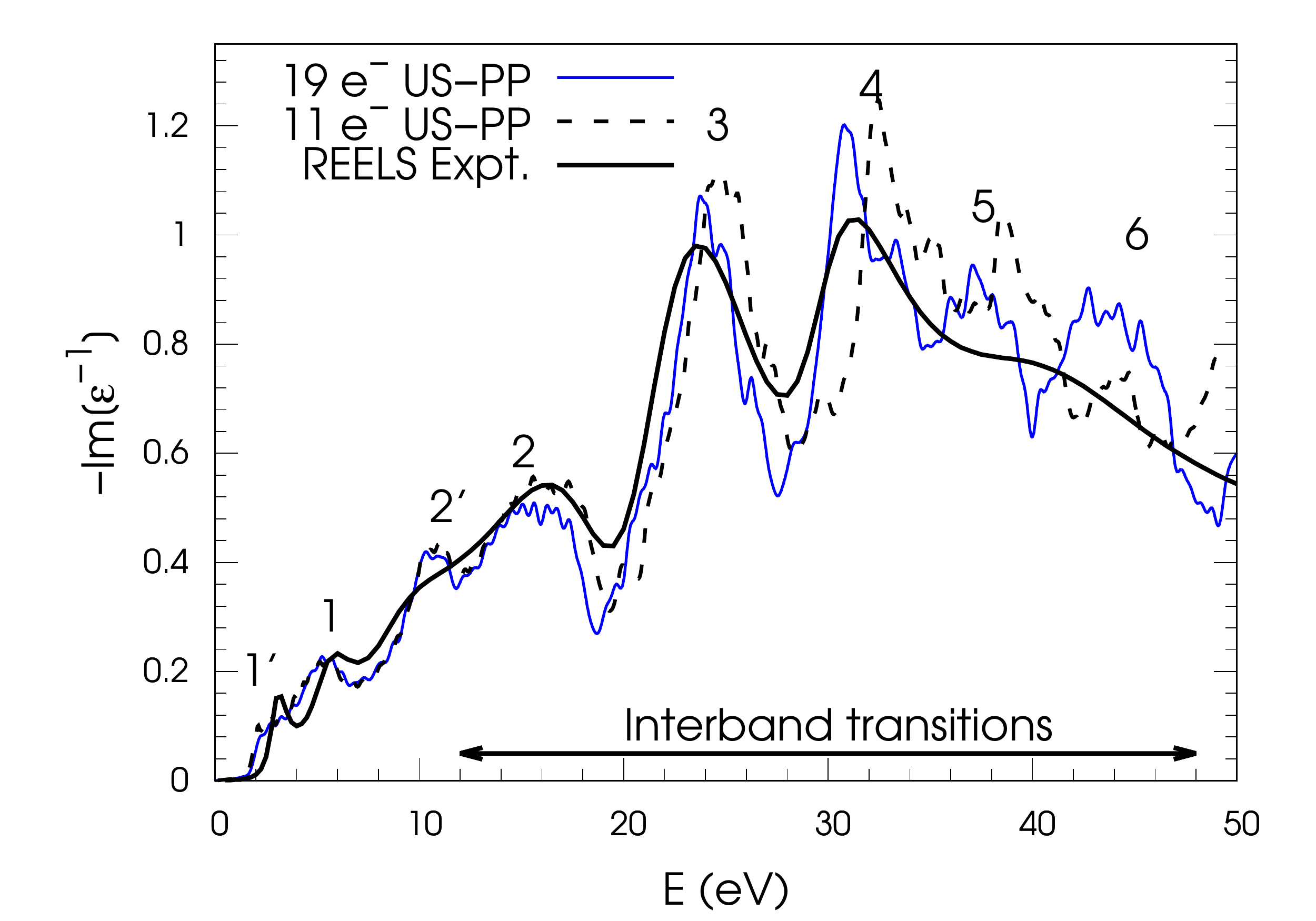}
  \caption{\label{fig:bulk} EEL spectrum of bulk Au computed with US-PP and $11$ valence electrons (black dashed line) or $19$ valence electrons (blue solid line) for $|\mathbf{q}|\approx 0.03$~\AA$^{-1}$ in the~[100] direction. Solid black line: from REELS experiment~\cite{Werner:2009} at \textbf{q}=\textbf{0}. Peak labels as in fig.~\protect\ref{fig:bulk_NC-PP}. 
Computations performed in TDDFPT-LDA.}  
\end{figure}

In fig.~\ref{fig:bulk} we present the loss functions of bulk Au obtained in TDDFPT-LDA, with
US-PPs having $19$ and $11$ electrons in the valence region (resp. solid-blue and dashed-black lines), 
and compare them with the loss function obtained in the reflection-EELS (REELS) 
experiment of ref.~\onlinecite{Werner:2009} (solid black line). We note that the difference between the EEL spectra computed with the same parameterization of US-PP but different functionals, GGA and LDA, at the same lattice parameter, look very similar (resp. figs.~\ref{fig:bulk_NC-PP}~and~\ref{fig:bulk}). 

%\Vchange{The US-PPs with 11 or 19 electrons in the valence region
%are in close-to-perfect agreement with each other up to $15$~eV (contribution 1$'$ and peaks $1$ and 2$'$).
% This validates the use of the US-PP with 11 electrons to study low energy excitations in  large systems made of gold atoms, and confirms an anterior work%~\cite{Gurtubay:2001}
%.
%Beyond 15~eV however, the US-PP with $11$ electrons overestimates the peak energy with respect to the 19 electron US-PP and the FP-LAPW results. This is due to missing contributions from the $5s$ and $5p$ semicore states (SSs) when only 11 electrons are accounted for in the valence region. }{}

Also we stress that according to our findings the 19 electrons US-PP must be used on the extended energy range (up to 50~eV in this work), while the 11 electrons US-PP is suitable for studies up to 15~eV only (see appendix~\ref{app:no_semicore} for more discussion about the semicore $5s$ and $5p$ states), as can be seen in Fig.~\ref{fig:bulk} when comparing with the experimental spectrum. Therefore, in the rest of this paper we present the results obtained with the 19 electrons US-PP.

\begin{table*}[t]
\centering
\begin{tabular}{llllllll}
\hline\hline
peak       &   \multicolumn{2}{c}{ This work }             &   \multicolumn{3}{c}{Ref.~\protect\onlinecite{Alkauskas:2013}} & Expt.~\protect\onlinecite{Werner:2009} & Expt.~\protect\onlinecite{Werner:2008} \\
           &    $\omega$ (eV)& Origin &   \multicolumn{2}{c}{$\omega$ (eV)} &   Origin                 &         \multicolumn{2}{c}{$\omega$ (eV)}   \\
 \#        &    19e US-PP             &                       &    FP-LAPW   &   $GW$               &                          &         \multicolumn{2}{c}{REELS}           \\
\hline
1$'$       &   2.2      & Mixed excitation      &  2.2  & 2.65 & Weak plasmon-like peak     & 3.25                     & 2.5               \\
1          &   5.1      & $5d$ plasmon          & 5.3   & 5.6  & Plasmon-type               & 6.0                      & 5.9   \\
2$'$       &   10.2     & Mainly-$6s$ plasmon   & 10.5  & 11.0 &                            & 10.2\footnotemark[1]     & 11.9\footnotemark[1]   \\
2          &   15.5     & IT                    & 15.4  & -    & IT                         & 16.3                     & 15.8 \\
3          &   23.8     & IT                    & 24.0  & -    & IT                         & 23.6                     & 23.6 \\
4          &   30.8     & IT                    & 31.1  & -    & IT                         & 31.2                     & 31.5 \\
5          &   36.9     & IT                    & 37.5  & -    & IT                         & -                        & 39.5 \\
6          &   43.5     & IT                    & 43.3  & -    & IT                         & -                        & 44.0 \\
\hline\hline
\footnotetext[1]{Appears as a shoulder in the spectrum.}
\end{tabular}
\caption{\label{tab:bulk_eels} Bulk Au. Energy of the peak positions in the loss function shown in fig.~\ref{fig:bulk} for the 19-electron US-PP in LDA. Peaks are labeled according to
ref.~\onlinecite{Alkauskas:2013} and we have added the labels
1$'$, 2$'$. 
``IT" stands for ``interband transition" and, by ``mixed excitation", we mean IT modified
by the intraband (Drude) contribution. ITs are present everywhere between 2.2 eV and 50 eV as a background contribution. 
}
\end{table*}

Experimental and theoretical peak positions are summarized in table~\ref{tab:bulk_eels}. 
A slight inaccuracy w.r.t. experiment comes from our neglect of spin-orbit coupling. 
Peak positions obtained 
with the $19$-electron US-PP are in good agreement with ref.~\onlinecite{Alkauskas:2013}, with differences between $-0.6$~eV and $0.2$~eV. With regard to experiment~\cite{Werner:2009}, peaks~$1$ and $2$ show differences of $-0.9$~eV and $-0.8$~eV, respectively, while in comparison with the experimental results of ref.~\onlinecite{Werner:2008}, peaks~$1$ and $2$ show differences of $-0.8$~eV and $-0.3$~eV, respectively.

Part of the inaccuracy comes from the description of $5d$ bands of Au in LDA~\cite{Rangel:2012,Alkauskas:2013} that leads to the 
redshift of the interband transition onset by approximately $0.5$~eV. 
As can be seen from table~\ref{tab:bulk_eels}, the redshift was only partially corrected by the approximate $GW$ calculation of ref.~\onlinecite{Alkauskas:2013} (table~\ref{tab:bulk_eels}). 

Finally, the contribution 1$'$ cannot be singled out as a peak, 
%\Vchange{and cannot be considered as a plasmon,}{} 
in contrast with experiment (fig.~\ref{fig:bulk}). This point is further discussed below.

\subsection{Origin and dispersion of the peaks}

\subsubsection{Interband transitions}

We found the presence of ITs between 2.2~eV and 50~eV, \textit{i.e.} the highest energy for which the EEL spectrum has been computed 
(fig.~\ref{fig:eps_bulk}, center panel, dotted line). 
Interband transitions are characterized in gold by a very weak dispersion above 12~eV, and peaks 2-6 are attributed to ``pure" interband transitions
(fig.~\ref{fig:eps_bulk_q}, bottom panel).  
Below 12~eV, ITs are dispersing, 
which is the fingerprint of their mixing with plasmon excitations, as will be discussed in section~\ref{subsubsec:mixed_exc} (fig.~\ref{fig:eps_bulk_q}, top panel).
The attribution of peaks beyond 15~eV is in agreement with ref.~\onlinecite{Alkauskas:2013}.

\subsubsection{Plasmon excitations} 

The energy of a collective excitation like a plasmon cannot be deduced directly from the band structure. 
Bulk plasmon energies are coming from the zeroes of the real part of the dielectric function, with a positive slope of the real part of $\epsilon$. The crossing of the zero energy axis 
strictly implies the existence of a self-sustaining oscillation. 

\begin{figure}[h!] 
    \centering
   \includegraphics[width=2.15in, angle=-90]{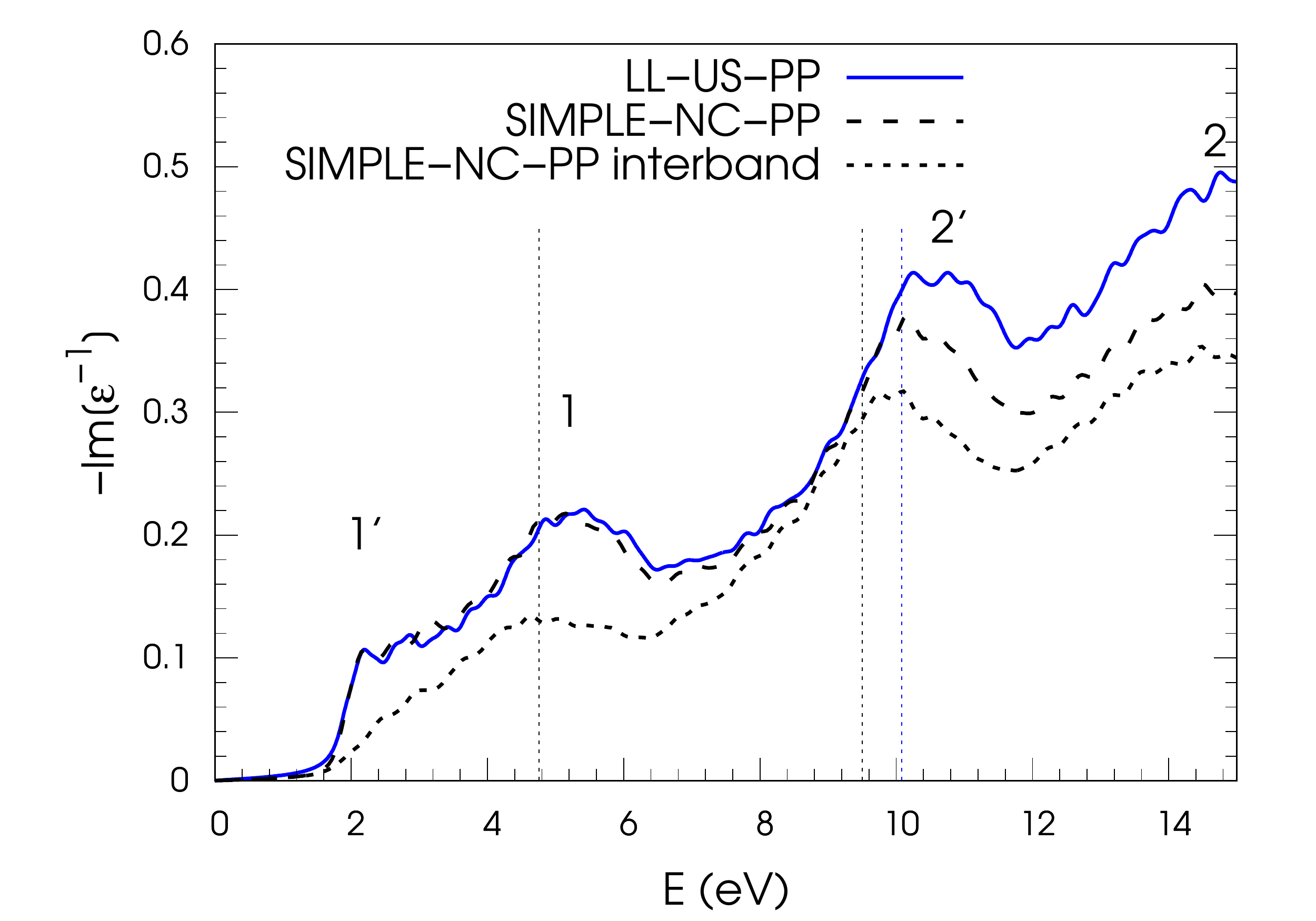}
   \includegraphics[width=2.15in, angle=-90]{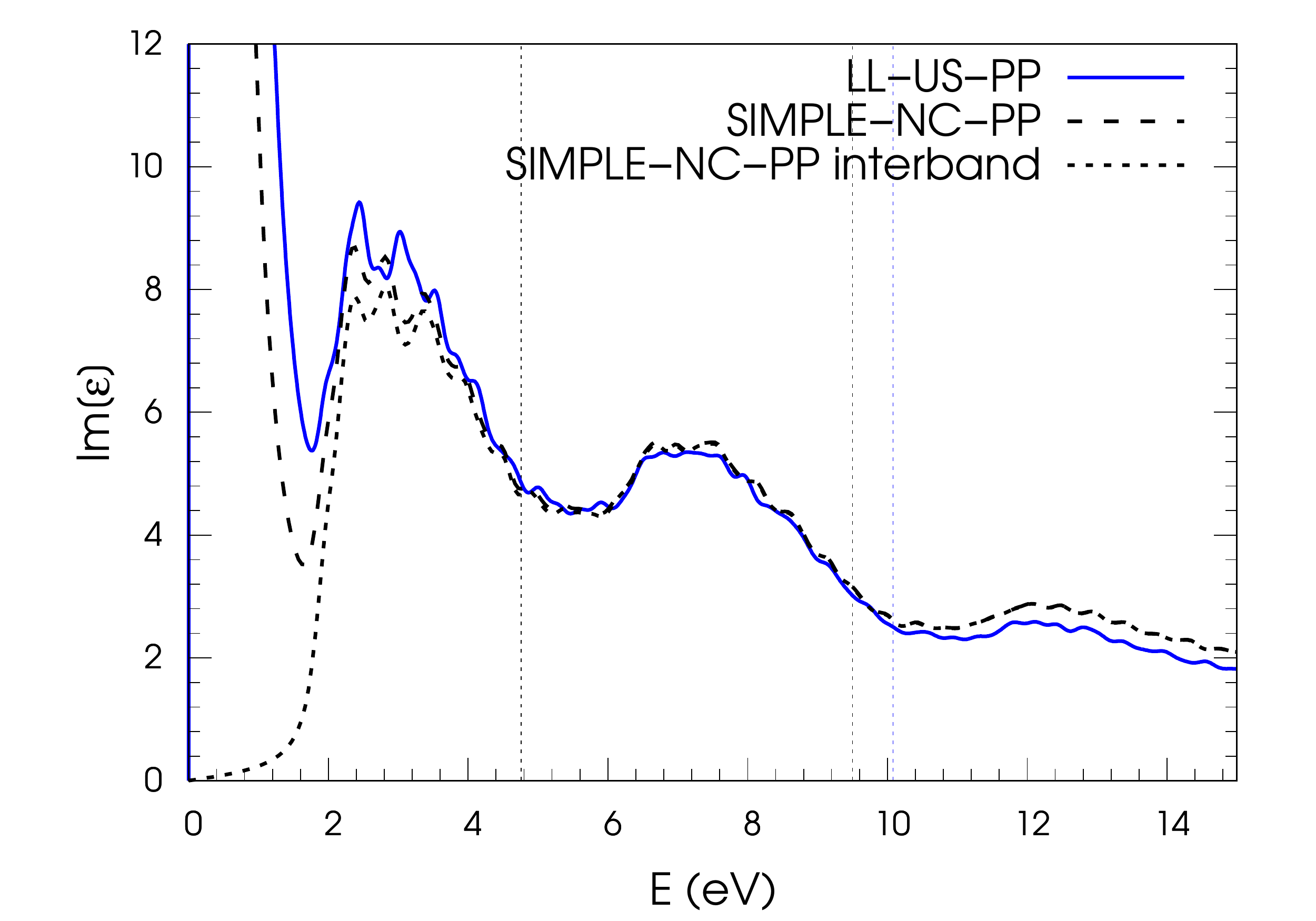}
   \includegraphics[width=2.15in, angle=-90]{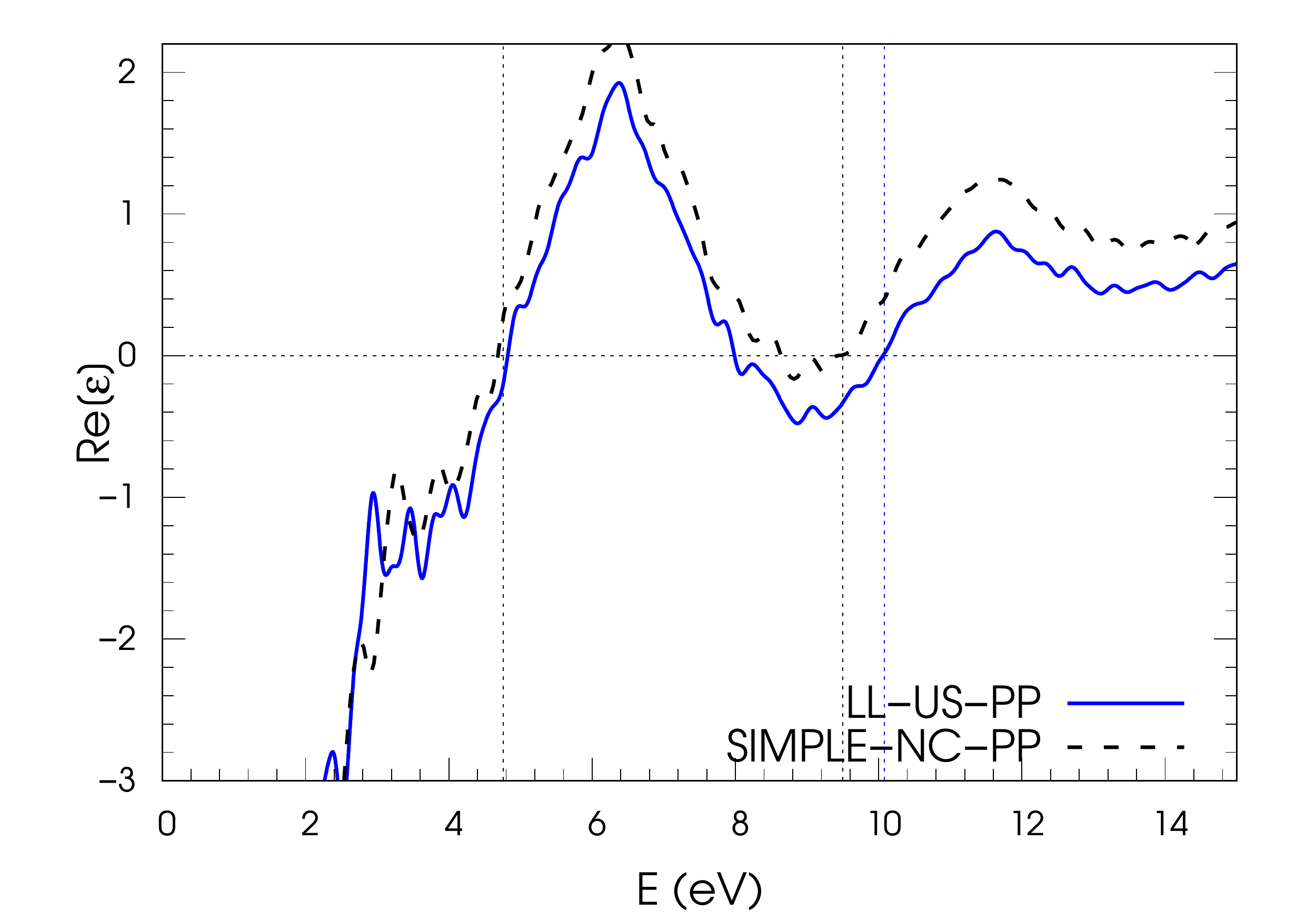}
  \caption{\label{fig:eps_bulk} Dielectric function (DF) and loss function (LF) of bulk Au with 19 electron PPs. Real part (bottom panel) and imaginary part (center panel) of the DF and LF (top panel) up to 15~eV for $|\mathbf{q}|\approx 0.03$~\AA$^{-1}$ in the~[100] direction. Solid lines: Liouville-Lanczos's (LL) method as implemented in the \texttt{turboEELS} code. Dashed lines: \textsc{Simple} code at \textbf{q}=\textbf{0} with both interband and intraband contributions. Dotted line: interband contributions only to the imaginary part of the DF at \textbf{q}=\textbf{0}. The vertical bars show the zeroes of the  real part (all panels).
Slights differences that are observed in the imaginary part and real part of the DF below 10~eV when comparing the calculation at \textbf{q}=\textbf{0} and the calculation at 
$|\mathbf{q}|\approx 0.03$~\AA$^{-1}$ comes from the diverging behavior of the intraband contribution at \textbf{q}=\textbf{0}. Differences that are observed above 10~eV  are due to the neglect of local fields in the calculation with the \textsc{Simple} code at \textbf{q}=\textbf{0}. Computations in TDDFPT-LDA.  
}
\end{figure}

In gold, there are two such oscillation frequencies, as the real part of the dielectric function crosses the zero energy axis at 4.8~eV 
and at 10.05~eV (fig.~\ref{fig:eps_bulk}, bottom panel). 
Therefore peaks 1 and 2$'$ of the loss function are unambiguously attributed to plasmons at 5.1~eV and 10.2~eV, respectively (fig.~\ref{fig:eps_bulk}, top~panel). The difference in frequency between the zeroes of the real part, and the positions of the two plasmon peaks, is explained  
below in the present section. The attribution of the peak at 5.1~eV to a plasmonic excitation is in agreement with ref.~\onlinecite{Alkauskas:2013}.

%\Vchange{More specifically, peak~ 2$'$, which has not been discussed in the literature so far, is attributed to the plasmon of the electron gas coming mainly form the $6s$ electrons. 
%Peak~$2^\prime$ was also found in the calculated spectrum ref.~\onlinecite{Alkauskas:2013} (see fig.~3 in the Supplemental Material. 
%This excitation shows up as a shoulder in the experimental spectrum (fig.~\ref{fig:bulk}). } {}

\begin{figure}[t]
    \centering
  \includegraphics[width=2.5in, angle=-90]{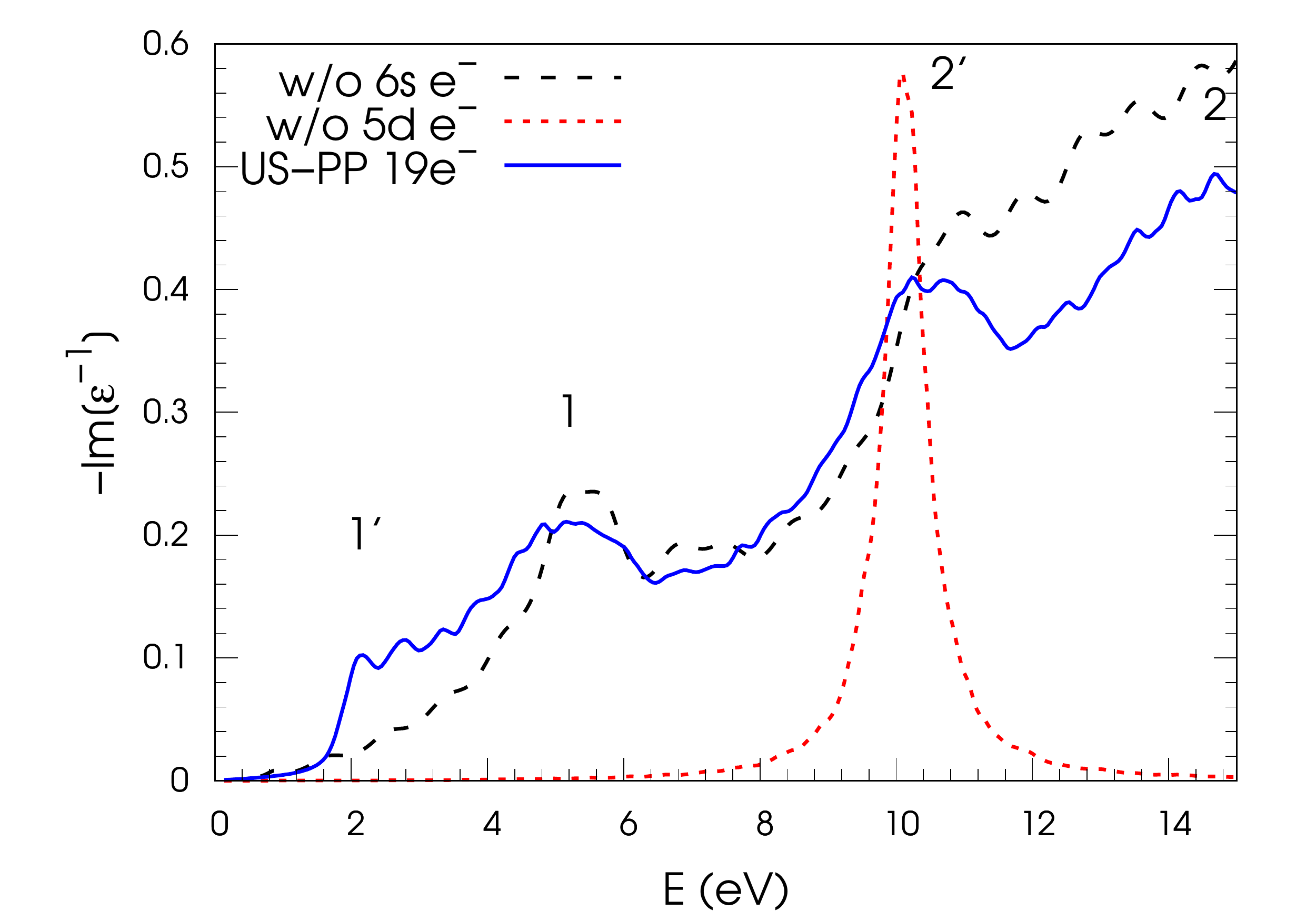}
  \caption{\label{fig:5d_6s} Toy-model EEL spectra computed using modified NC-PP pseudopotentials: without the $6s$ electron (black dashed line) or without $5d$ electrons (red dashed line). 
They are compared to the EEL spectrum of the 19-electron US-PP that contains both $5d$ and $6s$ electrons (blue solid line, same as in fig.~\ref{fig:bulk}). In the legend, ``w/o" stands for without. Results computed using the LL method. Computation performed within TDDFPT-LDA. 
}
\end{figure}

On the other hand, peak~ 2$'$, which is also found in the calculated spectrum of ref.~\onlinecite{Alkauskas:2013} (table~\ref{tab:bulk_eels}), has not been discussed so far. 
Thanks to the two toy-model PP calculations with only $6s$ electrons and with only $5d$ electrons, we unambiguously conclude that the plasmon peak $2^\prime$ is coming mainly from the collective excitation of $6s$ electrons. 
Indeed, in fig.~\ref{fig:5d_6s}, the spectrum computed with a toy-model containing only the $6s$ electron is represented by a single plasmon at 10.05~eV (red dotted line). There is hardly any peak in the calculation with only  $5d$ electrons and no $6s$ ones, while all of the calculations with more than 10 electrons consistently show a peak at 10.2~eV. 

On the other hand, peak~1 is attributed to a plasmon oscillation from $5d$ electrons. Indeed, in fig.~\ref{fig:5d_6s}, the spectrum computed with the toy model PP without the $6s$ electron, containing only $5d$ electrons, has a well pronounced peak at 5.1~eV (black dashed line), in close-to-perfect agreement with the plasmon positions reported in ref.~\onlinecite{Alkauskas:2013} (table~\ref{tab:bulk_eels}). 
% for both the 11-electron and 19-electron pseudopotentials. 
Thus, electrons in gold behave as two quasi-separate electron gasses, each one oscillating with its own frequency.   

However, as seen above, in our calculations, the two plasmons are intimately influenced by the interband transitions contained in $\mathrm{Im}(\epsilon)$ around the respective plasmon frequencies. 
This is the reason why  the zeros of the real part of the dielectric function (resp. 4.8~eV
and at 10.05~eV) and the main peaks of the energy loss function (5.1~eV and 10.2~eV) do not exactly coincide: the imaginary part is not minimal 
at the position of the zeros of the real part in the calculations, and the plasmon positions are blue-shifted by 0.3~eV and approximately 0.15~eV w.r.t. the zeros of the real part of the dielectric function. We point out that while crystal local fields have no effect on the zero energy crossing at 4.8~eV~coming from the $5d$ electron gas~(fig.~\ref{fig:eps_bulk}, bottom~panel, dotted vertical line on the left-hand side), they have an important effect at the 10.05~eV~(fig.~\ref{fig:eps_bulk}, bottom~panel, black and blue and dotted vertical lines on the right-hand side)~\cite{eMotornyi:Note:2019e}.  
We also note that in the experimental data reported around 10~eV, the real part of the dielectric function is found to be positive~\cite{Weaver:2015}. 

Finally, inspection of our calculations does not allow us to attribute a bulk plasmonic origin to the peak at 3.25~eV~\cite{Werner:2009} or 2.5~eV~\cite{Werner:2008}  observed in the REELS experiments (table~\ref{tab:bulk_eels}). 
%\Vchange{In fact, this peak is not observed in loss functions derived from optically measured dielectric functions~\cite{Olmon:2012,Babar:2015}.}{}
This point is discussed in more detail in the next section.

\subsubsection{Mixed excitations} 
\label{subsubsec:mixed_exc}

In this section we discuss the remaining peaks in the EEL spectra of bulk Au. There is no clear peak that can be singled out near 2~eV in the loss function (contribution~1$'$, fig.~\ref{fig:eps_bulk}, top panel) nor any zero of the real part of the dielectric function near 2~eV in the scalar-relativistic calculation (bottom panel). 
Thus contributions in the loss function between $\approx$~2 and $\approx$~4~eV are due to $5d \rightarrow 6s$ interband contributions. We note however that, in this energy interval,
the interband contributions are modified by the intraband component of the excitation coming from the presence of a plasmon at 4.8~eV in the real part of the dielectric function. This can be numerically checked by inserting in eq.~\eqref{eq:loss} $Re(\epsilon_B)$ of the total dielectric function (both intraband and interband) and $Im(\epsilon_B)$ containing only the interband contribution (not shown). Thus contribution~1$'$ near 2.2~eV is attributed to mixed excitations, and has no plasmonic origin (table~\ref{tab:bulk_eels}). 

On the other hand, the interpretation of the  contribution near 2.2~eV in previous works\cite{Alkauskas:2013} was made in analogy with the plasmon in silver. 
In silver, there is a plasmon at 3.8~eV whose position is due to the shift of the mainly-$s$ plasmon at 9.7~eV, caused by the presence of interband transitions~\cite{Alkauskas:2013}.
By analogy, in gold, it was thought that a very weak plasmon-like peak was developing at 2.65~eV when calculations were performed with methods beyond DFT, e.g. with the approximate $GW$ 
calculations (ref.~\onlinecite{Alkauskas:2013}, supplemental material). 
In gold however, we find the well-defined $s$ plasmon at 
10.2~eV (see previous section), and a related peak is also found in the $GW$ calculations at 11.0~eV (table~\ref{tab:bulk_eels}). 
%\Vchange{There is no any near-to-zero-crossing of the real part of the dielectric function around 2.2~eV. }{} 
%\Vchange{Instead, we find a wealth of $5d \rightarrow 6s$ interband contributions, over a broad energy band, near 2.2~eV.}
Consequently, the contribution near 2.2~eV, in bulk gold, is probably solely made of a wealth of $5d \rightarrow 6s$ interband contributions. 

Moreover, it should be noted that, from the experimental side, extra complications come from the fact that,  at a similar energy, several contributions show up~\cite{Werner:2008}. In particular, 
the deconvolution of the bulk and surface contributions from the experimental total spectra is very sensitive to many details. This makes it hard to determine precisely the exact position of the contribution $1^\prime$ (table~\ref{tab:bulk_eels}, last two columns on the right-hand side).

\begin{figure}[t]
    \centering
   \includegraphics[width=2.5in, angle=-90]{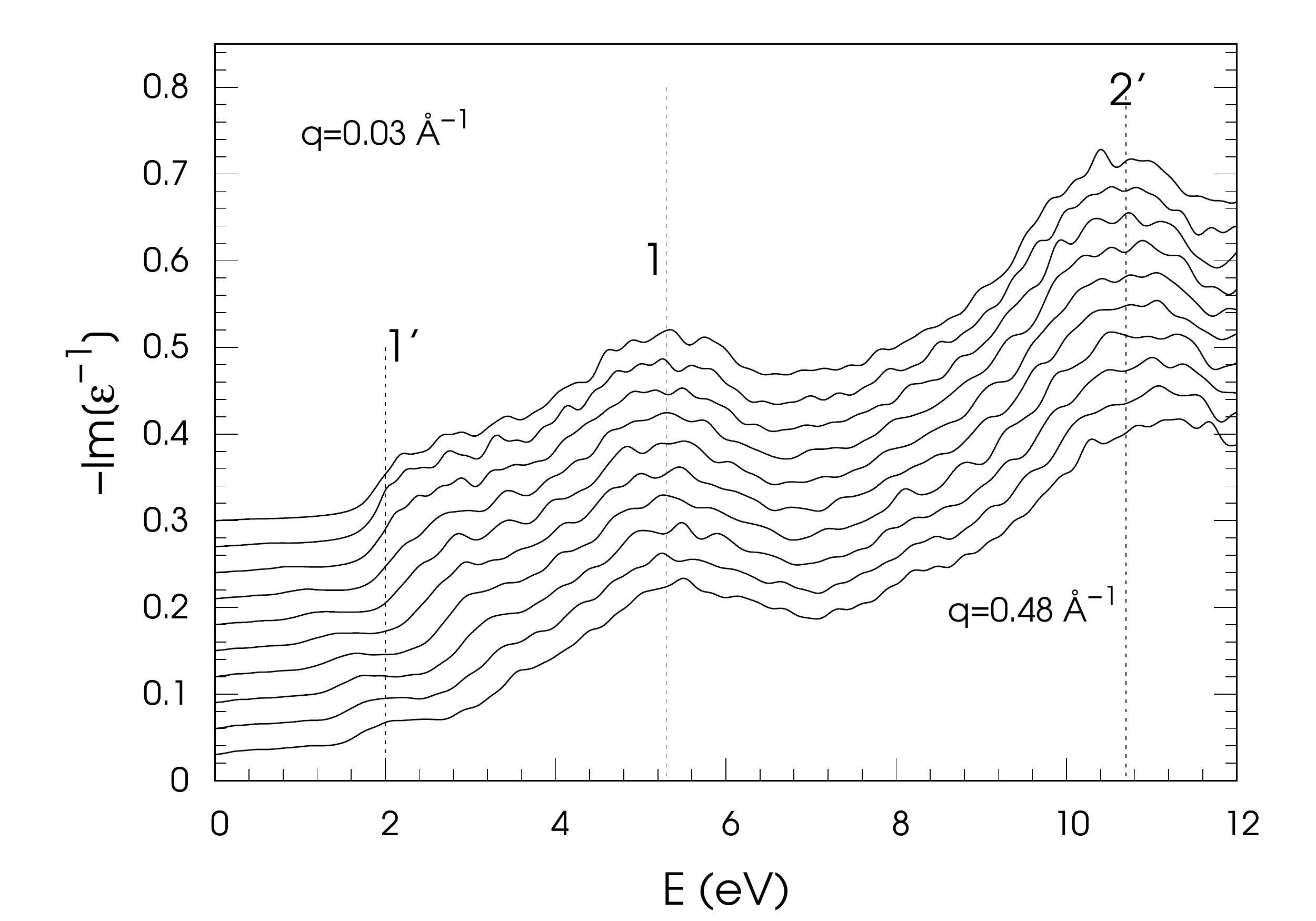}
   \includegraphics[width=2.5in, angle=-90]{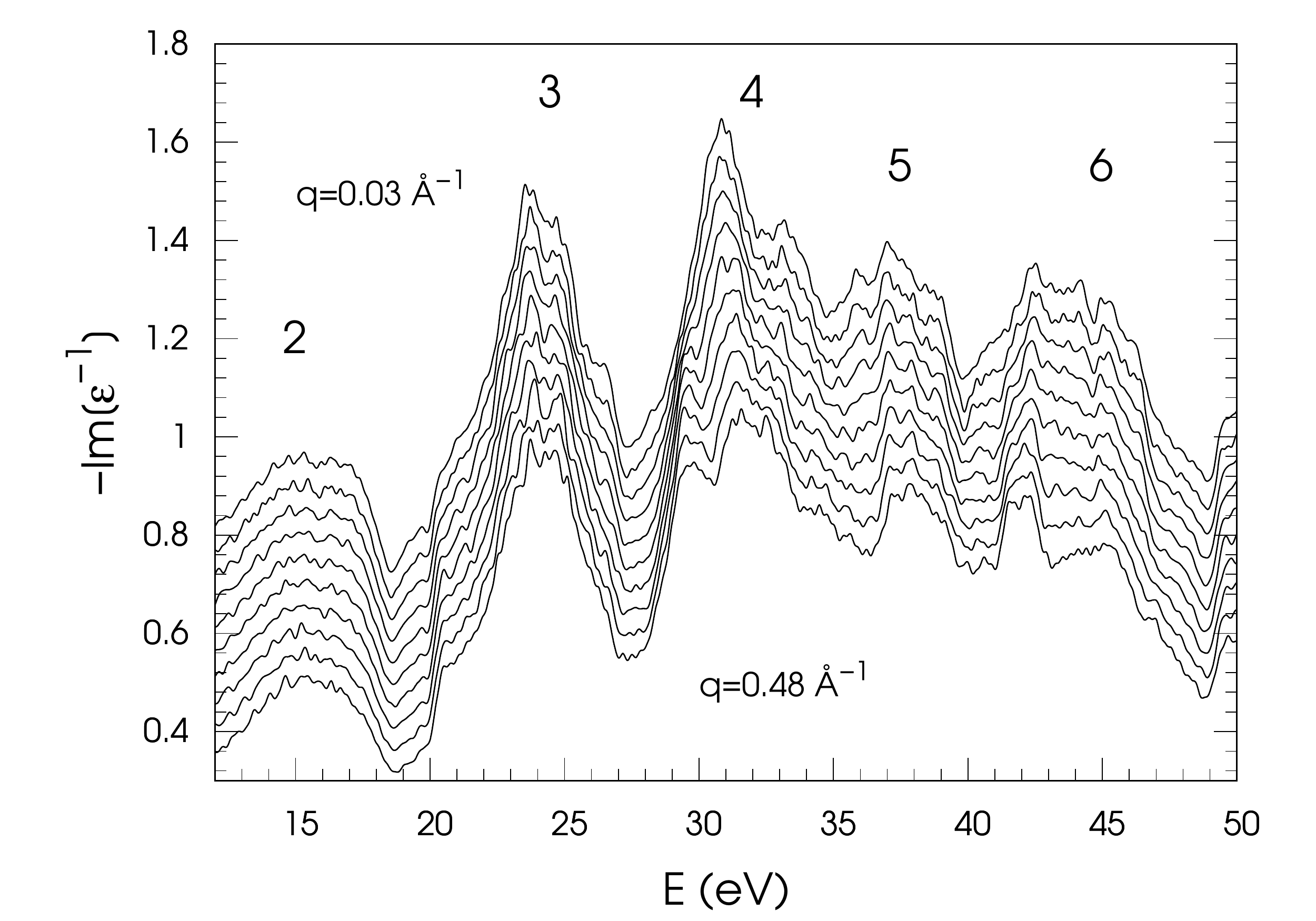}
  \caption{\label{fig:eps_bulk_q} Loss function of bulk Au with the 19 electron US-PP.
Loss function as a function of energy and transferred momentum $q$ in the~[100] direction (black solid lines).
Energy below 12~eV (top) and above 12~eV (bottom panel). The thin dotted lines in the top panel are guides for the eyes to highlight the peak dispersion. 
Results computed with  the LL method within TDDFPT-LDA. 
}
\end{figure}

Finally, the dispersion of the contribution~1$'$ is reported in fig.~\ref{fig:eps_bulk_q} (top panel). Indeed, as the loss function due to interband transitions is modified by the intraband contribution to the real part of the dielectric function, it shows some dispersion. This is the case for the contribution~1$'$ in gold, as well as for the peak at 5~eV in bulk bismuth~\cite{Timrov:2017}. This is general to materials in which there are interband transitions below the (not too far) plasmon energy: 
interband transitions then show some dispersion.

\section{Conclusion}
\label{sec:conclusion}  
In conclusion, in the present paper we have demonstrated how the Liouville-Lanczos and the Sternheimer approaches 
to the TDDFPT calculations of electron energy loss spectra and inelastic X-ray scattering cross-sections can be generalized to Vanderbilt's US-PPs.
We have  shown, on the example of bulk Au, that results obtained with 
the LL method and US-PPs agree with other TDDFT studies. 
%\Vchange{In particular, we have discussed the effect of the inclusion of semicore electrons in the valence region of the pseudopotential.}{}

We have analyzed in detail the origin of various peaks in the EEL spectra of bulk Au.
We have found that the signature of plasmons in EEL spectra of bulk Au revealed by our study shows the richness of the physics underlying the various contributions to the density fluctuation in gold. 
In particular, we have attributed peaks at 5.1~eV and 10.2~eV to plasmon excitations coming from $5d$ and $6s$ electron gasses, respectively. 
We have defined as a mixed excitation a contribution from interband transitions that lies below the plasmon frequencies and is modified by the real part of the intraband contribution,
and given a way to characterize numerically a mixed excitation in the calculations.
We have then identified the contributions between $\approx$~2.2~eV and $\approx$~4~eV  as mixed excitations. 
We have concluded that 
%\Vchange{our calculations for bulk Au}
alone, contributions of bulk Au cannot explain the presence of the well-defined peak at 
%\Vchange{3.25~eV}
low energy observed in the REELS experiments. 
%\Vchange{, and our conclusion is consistent with the the loss function deduced from optical data, which do not show a peak at 3.25~eV. }{}
Finally we have shown the dispersion of plasmons and mixed excitation, and the very weak dispersion of pure interband transitions at energies above the plasmon frequencies. 

\section{Acknowledgments}

Results have been obtained with the \texttt{turboEELS}   
and  \textsc{SIMPLE} codes of the   \textsc{Quantum ESPRESSO}~\cite{Giannozzi:2017} project, and with the \texttt{Thermo{\_}pw}~\cite{Dalcorso:2019} package. Computer time has been granted by the Partnership for Advanced Computing in Europe (PRACE Project No. 2010PA3750), by the national centers GENCI-CINES and GENCI-TGCC (Project 2210), and by \'Ecole Polytechnique through the LLR-LSI project.

O.M. acknowledges support from the doctoral school INTERFACES of \'Ecole Polytechnique, and from the EU-MaX project, and fruitful discussions with M.~Raynaud at an early stage of the project. This work was partly funded by the EU Commission through the MaX Centre of Excellence for Supercomputing Applications, grants no. 676598 and 824143.
I.T. acknowledges support from the Swiss National Science Foundation (SNSF), through grant 200021-179138, and its National Centre of Competence in Research (NCCR) MARVEL.  ADC acknowledges also
support from the SISSA ITCS and its Linux cluster.

\appendix
\section{Inverse overlap operator for periodic solids}
\label{app:Sm1}

In the US-PPs case, the inverse of the overlap operator $\hat{S}$ needs to be introduced in the 
Liouville-Lanczos + NC-PPs algorithm of the TDDFPT equations. 
It was introduced in 
ref.~\onlinecite{Walker:2007} as:
\begin{equation}
\hat{S}^{-1} = 1 + \sum_{IJ,mn} \lambda^{IJ}_{mn} |\beta^I_m\rangle\langle
\beta^J_n|,
\end{equation}
where the sum over $I$ and $J$ runs on the atoms of the system and
the sum over $m$ and $n$ runs over the projectors of each atom. 
In ref.~\onlinecite{Walker:2007}, the coefficients $\lambda^{IJ}_{mn}$
are obtained from the condition $\hat{S} \hat{S}^{-1}= 1$ which gives the linear
systems:
\begin{equation}
\sum_{Im} \left(\delta_{rm} \delta_{NI} + \sum_i q^N_{ri} \, B^{NI}_{im} \right)
\lambda^{IJ}_{mn} = -q^N_{rn} \delta_{NJ},
\label{ls}
\end{equation}
where $B^{NI}_{im}=\langle \beta^N_i | \beta^I_m\rangle$ and $\delta_{NI}$ 
is the Kroenecker symbol. 
There are $N_B$ linear systems (for each $J$ and $n$) of size 
$N_B \times N_B$, where $N_B$ is the number of projectors in the system.

In a periodic solid we can write $I=\{l, s\}$, $J=\{l', s'\}$ and $N=\{l'',s''\}$ and 
use periodicity to remove the dependence on the unit cell index, obtaining
in this way $N_b$ linear systems of dimension $N_b \times N_b$, where $N_b$
is the number of projectors in one unit cell. Clearly different systems will 
be obtained for each {\bf k}-point in the Brillouin zone.
To compute the action of the operator $\hat{S}^{-1}$ on a Bloch function 
$\psi_{{\bf k},i}({\bf r})$  
it is convenient to define the coefficients:
\begin{equation}
\lambda^{{\bf k},ss'}_{mn} = \sum_{l'} e^{-i{\bf k}\cdot{\bf R}_l} \,
\lambda^{ls, l's'}_{mn} \, e^{i{\bf k}\cdot{\bf R}_{l'}} ,
\end{equation}
where $\lambda^{{\bf k},ss'}_{mn}$ are independent from
$l$ since $\lambda^{ls,l's'}_{mn}$ depend on ${\bf R}_l-{\bf R}_{l'}$.
Inserting this definition in the expression of $\hat{S}^{-1}$ we find:
\begin{eqnarray}
\sum_{IJ, mn} &\lambda&^{IJ}_{mn} |\beta^I_m\rangle \langle \beta^J_n |
\psi_{{\bf k},v}\rangle= \nonumber \\ \sum_{ss', mn} &\lambda&^{{\bf k},ss'}_{mn} 
\sum_l |\beta^{I}_m\rangle  
e^{i{\bf k}{\bf R}_l} \langle \beta^{s'}_n | \psi_{{\bf k},v}\rangle.
\end{eqnarray}
For each ${\bf k}$, the $N_b^2$ coefficients $\lambda^{{\bf k},ss'}_{mn}$,
are solutions of the $N_b$ linear systems (for each $s'$ and $n$):
\begin{equation}
\sum_{sm} \left(\delta_{rm} \delta_{s''s} + \sum_i q^{s''}_{ri} \,
B^{{\bf k}, s''s}_{im} \right)
\lambda^{{\bf k},ss'}_{mn} = - q^{s''}_{rn} \delta_{s''s'},
\end{equation}
where
\begin{equation}
B^{{\bf k}, s''s}_{im} = \sum_l e^{i{\bf k}\cdot{\bf R}_l} \int d^3r \, 
\beta^{\gamma(s'')*}_i({\bf r} - {\boldsymbol \tau}_{s''})
\beta^{\gamma(s)}_m({\bf r}-{\bf R}_l - {\boldsymbol \tau}_s) .
\label{bdef}
\end{equation}
which can be obtained by multiplying Eq.~\ref{ls} by $e^{i\mathbf{k} \cdot \mathbf{R}_{l'}}$
and adding on $l'$. 

Finally, the operator $\hat{S}^{-1}_{\mathbf{k}}$ that appear in Eqs.~\eqref{eq:EELS_D_super-op_us}, 
\eqref{eq:EELS_K_super-op_us}, and \eqref{eq:EELS_K_super-op_rhs} can be obtained 
from the relationship: 
\begin{equation}
\hat{S}^{-1} \psi_{{\bf k},i}({\bf r}) = e^{i\mathbf{k}\cdot\mathbf{r}} \hat{S}^{-1}_{\mathbf{k}}
u_{{\bf k},i}({\bf r}) , 
\end{equation}
and can be written as:
\begin{equation}
   \hat{S}^{-1}_\mathbf{k} = 1 + \sum_{ss',mn} \lambda^{{\bf k},ss'}_{mn} | \beta^{s,\mathbf{k}}_m \rangle \langle \beta^{s',\mathbf{k}}_n |.
\end{equation}
We benchmarked the implementation of the operator $\hat{S}^{-1}_\mathbf{k}$ in the LL method by comparing the final results with the results obtained using the method based on the solution of the Sternheimer equations, since the latter does not require $\hat{S}^{-1}_\mathbf{k}$.

\section{Effect of the semicore states}
\label{app:no_semicore}

In the present appendix we discuss the effect of semicore $5s$ and $5p$ states on the EEL spectrum of bulk Au. 

This is an important aspect, because in many studies semicore states are frozen in the core, and hence it is important to clarify in which energy range this approximation gives reliable results. In fact, in the case of Bi it was shown that by freezing the $5d$ semicore states in the core region, the plasmon peak position is strongly affected and the high-energy part of the spectrum is completely missing~\cite{Timrov:2017}. In a previous study of the electronic bandstructure of gold, inclusion of the semicore states was shown to have practically no effect on the DFT-level, but to be very important in the $GW$ calculations~\cite{Rangel:2012}. Here we determine the energy range on which the EEL spectra are accurate with $5s$ and $5p$ semicore states frozen in the core in the case of gold, and also we highlight at which energies the effect of $5s$ and $5p$ in the valence region is crucial.

In this work we used two types of US-PPs: the 11 electrons case (with $5s$ and $5p$ semicore states frozen in the core) and 19 electrons case (with $5s$ and $5p$ semicore states included in the valence region of the electronic configuration). We find that the US-PPs with 11 and 19 electrons are in close-to-perfect agreement with each other up to $15$~eV, i.e. for contribution 1$'$ and peaks $1$ and 2$'$ (see Fig.~\ref{fig:bulk} and Table~\ref{tab:bulk_eels_semicore}). This validates the use of the US-PP with 11 electrons to study low energy excitations in  large systems made of gold atoms, and confirms an anterior work~\cite{Gurtubay:2001}. 
However, for energies above 15~eV there are significant deviations in the peak positions and in their intensities (peaks 2 -- 6).

Moreover, the origins of the peaks are also left unchanged: at low energy all is exactly the same, while at higher energies all peaks come from the interband transitions (with missing interband contributions in the 11 electrons case due to missing initial states ($5s$ and $5p$) which are frozen in the core).

Therefore, we conclude that the 11 electrons US-PP can be safely used to describe low energy excitations, while the 19 electrons case is absolutely needed for investigations of extended energy portions of the EEL spectra. 

\begin{table}[h]
\centering
\begin{tabular}{lllllllll}
\hline\hline
Peak \#   & 1$'$             & 1            & 2$'$                 & 2    & 3    & 4    & 5    & 6          \\ \hline
11~elec.  & 2.2              & 5.3          & 10.5                 & 16.1 & 24.7 & 32.5 & 38.5 & 44.0  \\
19~elec.  & 2.2              & 5.1          & 10.2                 & 15.5 & 23.8 & 30.8 & 36.9 & 43.5 \\
\hline\hline
\end{tabular}
\caption{\label{tab:bulk_eels_semicore} Peak positions (in eV) in the EEL spectrum of bulk Au (see fig.~\ref{fig:bulk}) computed with two types of US-PPs, containing 11 electrons (without semicore states) and 19 electrons (with semicore states) in the valence.
}
\end{table}


\begin{thebibliography}{60}
\bibitem [{bMo()}]{bMotornyi:Note:2019b}%
  \BibitemOpen
  \href@noop {} {}\bibinfo {note} {The self-consistent cycle was ended when the
  estimated energy error was smaller than the threshold of 10$^{-12}$ for bulk
  Au.}\BibitemShut {Stop}%
\bibitem [{eMo()}]{eMotornyi:Note:2019e}%
  \BibitemOpen
  \href@noop {} {}\bibinfo {note} {We stress that also with the 19-electron
  NC-PP, the real part of the dielectric function shows two zero-energy
  crossings, and also in this case the crossing at high energy is more
  pronounced when crystal local field effects are accounted for:
  $\mathrm{Re}(\epsilon)$ goes down to about the value of~-0.4 while the
  crossing in the RPA without crystal local field effects,
  $\mathrm{Re}(\epsilon)$ goes down only to~-0.05.}\BibitemShut {Stop}%
\end{thebibliography}
\end{document}